%% file: main.tex
\documentclass[preprint,5p]{elsarticle}

\usepackage[english]{babel}
\usepackage[utf8]{inputenc}
\usepackage[T1]{fontenc}
\usepackage[pdftex]{hyperref}

\usepackage{amssymb}
\usepackage{amsmath}
\usepackage{amsthm}
\usepackage{amsfonts}
\usepackage{mathtools}
\usepackage{array}
\usepackage[linesnumbered, ruled, vlined]{algorithm2e}

\usepackage{multirow}
\usepackage{adjustbox} 
\usepackage{stfloats} 
\usepackage[caption=false]{subfig} 
\usepackage{booktabs} 
\usepackage{url}
\usepackage{wrapfig} 
\usepackage{graphicx}
\usepackage{textcomp}

\usepackage[colorinlistoftodos]{todonotes} 
\usepackage{xcolor} 
\def\BibTeX{{\rm B\kern-.05em{\sc i\kern-.025em b}\kern-.08em 
    T\kern-.1667em\lower.7ex\hbox{E}\kern-.125emX}}
\usepackage{colortbl} 
\usepackage{pifont} 
\usepackage[clock]{ifsym} 

\usepackage{pgfplots} 
\usepackage{tikz}
\usetikzlibrary{pgfplots.statistics, pgfplots.colorbrewer}
\usepgfplotslibrary{fillbetween} 
\usepgfplotslibrary{colorbrewer} 
\pgfplotsset{compat = 1.8, cycle list/Set3} 
\usetikzlibrary{} 
\usepackage{pgfplotstable}

\hypersetup{pdfborder=0 0 0}
\newcolumntype{L}[1]{>{\raggedright\arraybackslash}p{#1}}
\newcolumntype{C}[1]{>{\centering\arraybackslash}p{#1}}
\newcolumntype{R}[1]{>{\raggedleft\arraybackslash}p{#1}}
\newcommand{\colorcell}{ \cellcolor{gray!35} }

\newcommand{\definitionname}{Def.}

\newcommand{\sectionname}{Sec.}
\definecolor{customblue}{HTML}{BDE0FF}
\definecolor{customgray}{HTML}{E6E6E6}
\definecolor{customred}{HTML}{FFCCCC}

\theoremstyle{definition}
\newtheorem{defn}{Definition}[]

\makeatletter
\def\th@definition{
  \thm@notefont{}
  \normalfont 
}
\makeatother

\makeatletter
\pgfplotsset{
    boxplot prepared from table/.code={
        \def\tikz@plot@handler{\pgfplotsplothandlerboxplotprepared}%
        \pgfplotsset{
            /pgfplots/boxplot prepared from table/.cd,
            #1,
        }
    },
    /pgfplots/boxplot prepared from table/.cd,
        table/.code={\pgfplotstablecopy{#1}\to\boxplot@datatable},
        row/.initial=0,
        make style readable from table/.style={
            #1/.code={
                \pgfplotstablegetelem{\pgfkeysvalueof{/pgfplots/boxplot prepared from table/row}}{##1}\of\boxplot@datatable
                \pgfplotsset{boxplot/#1/.expand once={\pgfplotsretval}}
            }
        },
        make style readable from table=lower whisker,
        make style readable from table=upper whisker,
        make style readable from table=lower quartile,
        make style readable from table=upper quartile,
        make style readable from table=median,
        make style readable from table=average,
        make style readable from table=lower notch,
        make style readable from table=upper notch
}
\makeatother

\begin{document}

    \graphicspath{{figures/}} 

\input{frontmatter} 

\input{introduction}

\input{related-work}
    \input{approach}

\input{evaluation}

\input{conclusions}

\input{acknowledgements}

    \bibliographystyle{elsarticle-harv}
    \bibliography{references}

\end{document}

%% file: frontmatter.tex
\begin{frontmatter}

\title{Enhancing Business Process Simulation Models with Extraneous Activity Delays}

\author[ut]{David Chapela-Campa\corref{corr}}
\ead{david.chapela@ut.ee}

\author[ut]{Marlon Dumas}
\ead{marlon.dumas@ut.ee}

\cortext[corr]{Corresponding author}
\address[ut]{University of Tartu, Tartu, Estonia}

\begin{abstract}

Business Process Simulation (BPS) is a common approach to estimate the impact of changes to a business process on its performance measures. For example, it allows us to estimate what would be the cycle time of a process if we automated one of its activities, or if some resources become unavailable. The starting point of BPS is a business process model annotated with simulation parameters (a BPS model).
In traditional approaches, BPS models are manually designed by modeling specialists.
This approach is time-consuming and error-prone.
To address this shortcoming, several studies have proposed methods to automatically discover BPS models from event logs via process mining techniques.
However, current techniques in this space discover BPS models that only capture waiting times caused by resource contention or resource unavailability.
Oftentimes, a considerable portion of the waiting time in a business process corresponds to extraneous delays, e.g., a resource waits for the customer to return a phone call.
This article proposes a method that discovers extraneous delays from event logs of business process executions.
The proposed approach computes, for each pair of causally consecutive activity instances in the event log, the time when the target activity instance should theoretically have started, given the availability of the relevant resource.
Based on the difference between the theoretical and the actual start times, the approach estimates the distribution of extraneous delays, and it enhances the BPS model with timer events to capture these delays.
An empirical evaluation involving synthetic and real-life logs shows that the approach produces BPS models that better reflect the temporal dynamics of the process, relative to BPS models that do not capture extraneous delays.
\end{abstract}


\end{frontmatter}

%% file: introduction.tex
\section{Introduction\label{sec:introduction}}

Business Process Simulation (BPS) is an analysis technique that enables users to address ``what-if'' analysis questions, such as ``what would be the cycle time of the process if the rate of arrival of new cases doubles?'' or ``what if 10\% of the workforce becomes unavailable for an extended time period?''.
The starting point of BPS is a process model (e.g., in the Business Process Model and Notation -- BPMN)\footnote{\url{https://www.bpmn.org/}} enhanced with simulation parameters (a BPS model).
The usefulness of BPS hinges on the quality of the BPS model used as input.
Specifically, one expects that a simulation of a BPS model leads to a collection of simulated traces (herein a \emph{simulated log}), which resembles the traces observed in the real executions of the process.

Traditionally, BPS models are manually created by modeling experts. This task is time-consuming and error-prone~\cite{DBLP:journals/kais/MarusterB09}.
To tackle this shortcoming, several studies have advocated for the use of process mining techniques to automatically discover BPS models from business process event logs~\cite{DBLP:journals/peerj-cs/CamargoDR21,DBLP:journals/dss/CamargoDG20,DBLP:journals/bise/MartinDC16,DBLP:journals/is/RozinatMSA09}.
In this context, an \emph{event log} is a set of events recording a collection of activity instances pertaining to a process.
Each activity instance consists of an identifier of a process instance (trace), an activity label, the start and end timestamp of the activity instance, the resource who performed the activity, and possibly some additional attributes.
Given this input, automated BPS model discovery methods learn a BPS model that, when simulated, leads to simulated logs that replicate the real process, as recorded in the original log.


Current approaches for automated BPS model discovery generate BPS models under the assumption that all waiting times are caused either by unavailability of resources (e.g., resources being off-duty) or by resource contention (all resources who can perform an activity instance are busy performing other activity instances).
In practice, there are other sources of waiting times in business processes.
For example, Andrews et al.~\cite{DBLP:conf/pakdd/AndrewsW17} found that a large proportion of waiting time in insurance processes is not caused by the availability or capacity of resources, but by other factors, such as the resource working in another process, or the need for the customer to provide additional documents (where the receipt of those documents is not explicitly captured in the log).

\begin{figure}[t]
    \centering
    \includegraphics[width=0.98\columnwidth]{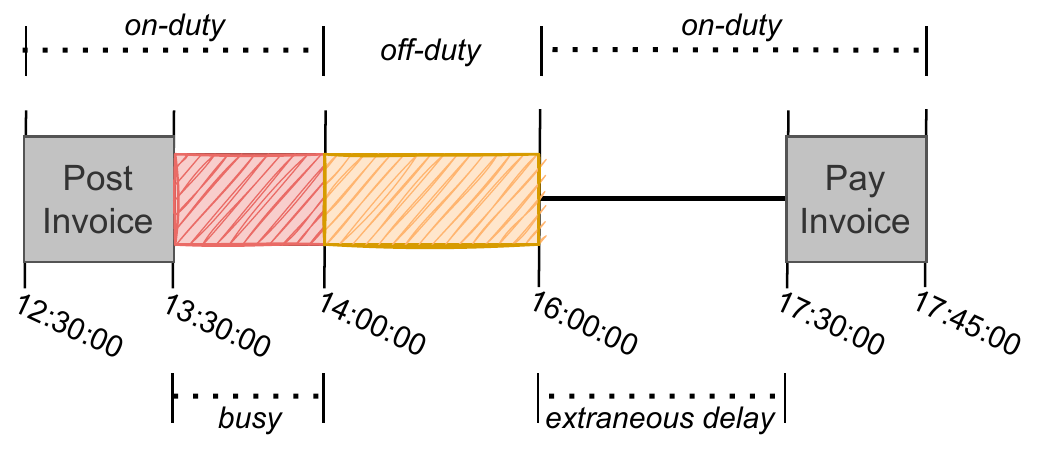}

    \caption{Example of an activity instance delayed due to resource contention, resource unavailability, and extraneous delays.}
    \label{fig:extraneous-delay-simple-example}
\end{figure}

This article is concerned with modeling \emph{extraneous delays} such as the above ones. In this context, an \emph{extraneous delay} is a delay in the start of an activity instance that is not attributable to resource unavailability (i.e., the resource being off-duty) nor attributable to resource contention (i.e., the resource being busy completing other activity instances of the same process). As an illustration, Figure~\ref{fig:extraneous-delay-simple-example} depicts a scenario where following the completion of an activity ``Post Invoice'', another activity ``Pay invoice'' is enabled. This latter activity does not start immediately, firstly because the resource who will perform it is busy performing some other activity instance (red box), and secondly, because this resource goes off-duty (e.g., evening time). Once the resource comes back on duty, an additional delay is observed until the ``Pay invoice'' activity starts. This latter delay is \emph{extraneous}, as it cannot be attributed either to resource contention or to unavailability.

This article studies the hypothesis that by taking into account extraneous waiting times during automated BPS model discovery, we obtain BPS models that better reflect the temporal performance of the process.
Specifically, the article addresses the following problem: given an event log $\mathcal{L}$ (wherein each activity instance has a resource, start time, and end time) and a BPS model $\mathcal{M}$ of a process, compute an enhanced BPS model $\mathcal{M}^{\prime}$ that incorporates waiting times caused by extraneous factors observed in the event log, such that $\mathcal{M}^{\prime}$ replicates the behavior recorded in $\mathcal{L}$ more accurately than $\mathcal{M}$.

To address this problem, we propose a configurable technique that enhances the BPS model by adding (duration-based) timer events to model the delays that cannot be explained by resource contention or unavailability.
One of the variants of the proposed technique computes, for each pair of causally consecutive activity instances in the event log, the time when the target activity instance should theoretically have started, given the availability of the resource that performed it.
Based on the difference between the theoretical and the actual start times, the approach estimates the extraneous activity delays.
We then propose two more variants that consider the first and the last time points when the target activity instance could theoretically have started.
The approach then fits a probability distribution function to model the discovered extraneous delays, and it enhances the BPS model with timer events that capture the delays modeled by the discovered distribution functions.

This article is an extended and revised version of a previous conference paper~\cite{DBLP:conf/icpm/Chapela-CampaD22}.
The article addresses the following limitations of the original technique presented in the conference paper:
\begin{itemize}
    \item It identifies waiting times due to resource unavailability due to working calendars, in addition to resource contention, and discards them from the extraneous delay computation.
    \item It considers the fact that delays may occur after an activity completes (ex-post delays), or before an activity starts (ex-ante delays), whereas the conference paper treats all extraneous delays as if they had occurred before the start of an activity. An example of an ex-post delay is a worker contacting a customer after recording a purchase order change, and blocking the execution flow until the customer is contacted. An ex-ante delay occurs, for example, when prior to triggering a shipment, a worker waits for oral confirmation of the shipment address.
    \item It considers the fact that between any two causally consecutive activity instances $\varepsilon_1$ and $\varepsilon_2$, the resource who performed $\varepsilon_2$ may have performed one or more other activity instances, which may ``eclipse'' parts of a (potential) extraneous delay between $\varepsilon_1$ and $\varepsilon_2$ (see \sectionname~\ref{subsubsec:eclipse-aware}).
\end{itemize}

The article reports on an evaluation that assesses: \textit{i)} the ability of the proposed technique to accurately discover extraneous delays known to be present in a log; \textit{ii)} the impact of modeling delays ex-post versus ex-ante; and \textit{iii)} the extent to which BPS models discovered using different variants of the proposed approach, more closely replicate the temporal dynamics in a log, relative to a baseline BPS model without extraneous delays.


The rest of the article is structured as follows.
\sectionname~\ref{sec:related-work} introduces basic notions of process mining and business process simulation, and gives an overview of prior related research. \sectionname~\ref{sec:approach} describes the proposed approach.
\sectionname~\ref{sec:evaluation} discusses the empirical evaluation, and \sectionname~\ref{sec:conclusions} draws conclusions and sketches future work.

%% file: related-work.tex
\section{Background and Related Work\label{sec:related-work}}

This section introduces basic notions of process mining related to the problem that is being addressed, and describes existent techniques related to the enhancement of BPS models.

\subsection{Event Log and Process Mining}

We consider a business process that involves a set of \emph{activities} $A$.
We denote each of these activities with $\alpha$.
An \emph{activity instance} $\varepsilon = (\varphi, \alpha, \tau_{s}, \tau_{e}, \rho)$ denotes an execution of the activity $\alpha$, where $\varphi$ identifies the \emph{process trace} (i.e., the execution of the process) to which this event belongs to, $\tau_{s}$ and $\tau_{e}$ denote, respectively, the instants in time in which this activity instance started and ended, and $\rho$ identifies the \emph{resource} that performed the event.
Accordingly, we write $\varphi(\varepsilon_{i})$, $\alpha(\varepsilon_{i})$, $\tau_{s}(\varepsilon_{i})$, $\tau_{e}(\varepsilon_{i})$, and $\rho(\varepsilon_{i})$ to denote, respectively, the process trace, the activity, the start time, the end time, and the resource associated with the activity instance $\varepsilon_{i}$.
We denote with $\omega(\varepsilon_{i})$ the \emph{waiting time} of $\varepsilon_{i}$, representing the time since $\varepsilon_{i}$ became available for processing, until its recorded start ($\tau_{s}(\varepsilon_{i})$).
The \emph{cycle time} of a trace (a.k.a.\ the trace duration) is the difference between the largest $\tau_{e}$ and the smallest $\tau_{s}$ of the activity instances observed in the trace.
An \emph{activity instance log} $\mathcal{L}$ is a collection of activity instances recording the information of the execution of a set of traces of a business process.
\tablename~\ref{tab:activity-instance-log-example} shown an example of 10 activity instances from an activity instance log.

It must be noted that, usually, the process execution information is recorded in event logs, containing the information of each activity instance as independent events, one representing its start, and another one representing its end.
Furthermore, there are cases in which event logs also store events recording other stages of the activity instances' lifecycle, e.g., schedule.
In this paper, we work with the concept of activity instance log to group the information of each activity instance in a single element.
The transformation from a typical event log to an activity instance log is straightforward, and it has already been described in other research \cite{DBLP:journals/is/MartinPM21}.
Nevertheless, we will use both terms: \textit{event log} and \textit{activity instance log}, to refer to the recorded process information used as starting point; and \textit{event} and \textit{activity instance}, to refer to the recording of an activity execution.

We use $\langle\tau_{i}, \tau_{j}\rangle$, such that $\tau_{i} < \tau_{j}$, to denote the time interval between $\tau_{i}$ and $\tau_{j}$.
In this way, the processing time of an activity instance $\varepsilon_{i}$ is denoted by $\langle\tau_s(\varepsilon_{i}), \tau_e(\varepsilon_{i})\rangle$.
We use $\tau_{k} \in \langle\tau_{i}, \tau_{j}\rangle$ to denote that the timestamp $\tau_{k}$ is contained in the interval $\langle\tau_{i}, \tau_{j}\rangle$, i.e., $\tau_{i} \leq \tau_{k}$ and $\tau_{k} \leq \tau_{j}$; and $\langle\tau_{i},\tau_{j}\rangle \models \langle\tau_{l},\tau_{m}\rangle$ to denote that the interval $\langle\tau_{i},\tau_{j}\rangle$ is contained in $\langle\tau_{l},\tau_{m}\rangle$, i.e., $\tau_{i} \geq \tau_{l}$ and $\tau_{j} \leq \tau_{m}$.
Finally, with $\langle\tau_{i},\tau_{j}\rangle \perp \langle\tau_{l},\tau_{m}\rangle$ we denote that both intervals (partially or fully) overlap, i.e., $\exists_{\langle\tau_{n},\tau_{o}\rangle \models \langle\tau_{i},\tau_{j}\rangle} \mid \langle\tau_{n},\tau_{o}\rangle \models \langle\tau_{l},\tau_{m}\rangle$.

\subsection{Business Process Simulation}

A process model is a diagrammatic representation of a process commonly used to model its control-flow dimension.
In this article, we consider process models represented in the Business Process Model and Notation (BPMN).
To model the control-flow of a process, a BPMN model consists of \textit{i)} a set of nodes representing the activities of the process; \textit{ii)} a set of gateways modelling the flow of the process, splitting or joining the flow through one or many paths; and \textit{iii)} a set of directed edges relating the activities and gateways between them.

However, a process model can also be extended with information about the execution of the process (e.g., the distribution of activities' processing times) or about the resources that can perform each specific activity.
Such enhanced process models are called business process simulation models.
Furthermore, in the BPMN formalism, other element types can be added to enrich the simulation, such as timer events (i.e., elements that force the process to wait an amount of time before continuing with the simulation).
When a BPS model is simulated it produces, based on the information it stores, an event log that replicates the process' behavior.

\begin{table*}[t]
    \centering
    \footnotesize

    \caption{Running example: fraction of an activity instance log with 10 activity instances storing the trace ($\varphi$), the executed activity ($\alpha$), the start ($\tau_{s}$) and end ($\tau_{e}$) times, and the resource ($\rho$).}
    \label{tab:activity-instance-log-example}

    \begin{tabular}{l l l l l}
    \toprule
         \multicolumn{1}{c}{\textbf{Trace}}  &  \multicolumn{1}{c}{\textbf{Activity}}   &  \multicolumn{1}{c}{\textbf{Start Time}}    &  \multicolumn{1}{c}{\textbf{End Time}}     &  \multicolumn{1}{c}{\textbf{Resource}}   \\ \toprule \toprule

        \multicolumn{5}{c}{$\vdots$}                                                          \\
        512   & Register invoice    &  03/11/2021 08:00:00  & 03/11/2021 08:31:11  & BoJack   \\
        512   & Post invoice        &  03/11/2021 08:31:11  & 03/11/2021 08:58:09  & Sarah    \\
        513   & Register invoice    &  03/11/2021 08:31:11  & 03/11/2021 09:02:51  & BoJack   \\
        512   & Notify acceptance   &  03/11/2021 09:00:00  & 03/11/2021 09:17:01  & Carolyn  \\
        514   & Register invoice    &  03/11/2021 09:02:51  & 03/11/2021 09:10:36  & BoJack   \\
        513   & Post invoice        &  03/11/2021 09:02:51  & 03/11/2021 09:35:50  & Sarah    \\
        513   & Notify acceptance   &  03/11/2021 09:17:01  & 03/11/2021 09:46:12  & Carolyn  \\
        514   & Notify acceptance   &  03/11/2021 09:46:12  & 03/11/2021 10:34:23  & Carolyn  \\
        514   & Post invoice        &  03/11/2021 11:00:00  & 03/11/2021 11:29:22  & Todd     \\
        512   & Pay invoice         &  03/11/2021 15:17:01  & 03/11/2021 15:27:45  & BoJack   \\
        513   & Pay invoice         &  03/11/2021 15:46:12  & 03/11/2021 15:57:43  & BoJack   \\
        514   & Pay invoice         &  04/11/2021 08:00:00  & 04/11/2021 08:31:02  & BoJack   \\
        \multicolumn{5}{c}{$\vdots$}                                                          \\ \bottomrule

    \end{tabular}
\end{table*}

\begin{figure}[t]
    \centering
    \includegraphics[width=0.98\columnwidth]{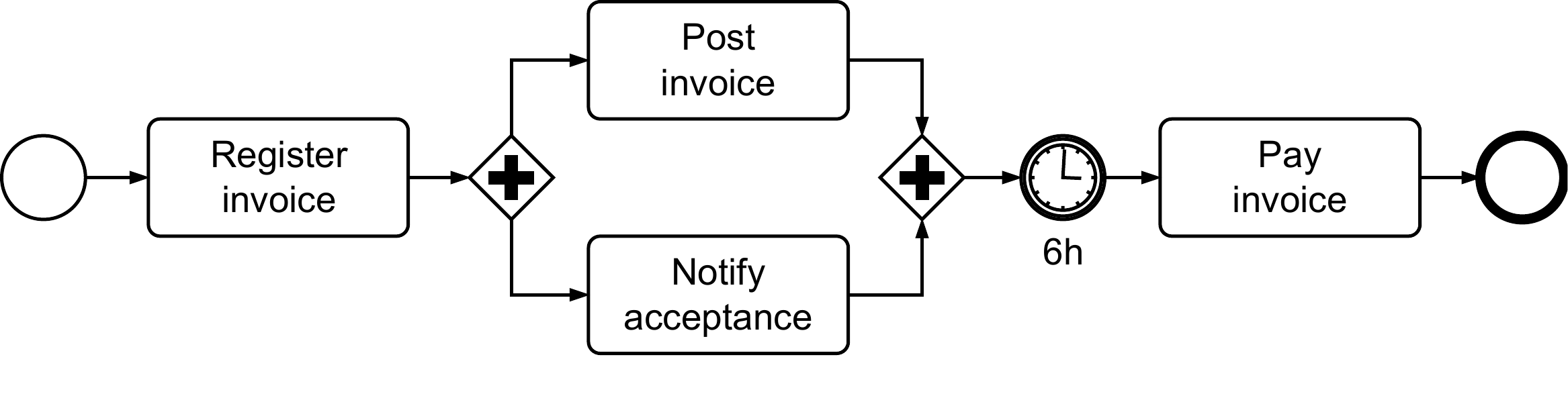}
    \caption{BPS model example formed by 4 activities, two AND gateways, and a timer event, corresponding the event log of \tablename~\ref{tab:activity-instance-log-example}.}
    \label{fig:bpmn-model-example}
\end{figure}

\figurename~\ref{fig:bpmn-model-example} shows an example of a BPS model corresponding to the event log example from \tablename~\ref{tab:activity-instance-log-example}, formed by 4 activities and one timer event.
As we can see in \tablename~\ref{tab:activity-instance-log-example}, the information about the waiting time is not explicitly present in the event log, but can be represented in the BPS model with a timer event.

\subsection{Related Work}

Since the apparition of the first approaches for automated BPS model discovery~\cite{DBLP:journals/is/RozinatMSA09}, many techniques have been presented focusing on the enhancement of BPS models to improve the accuracy and precision of the simulation.
Camargo et al. presented SIMOD~\cite{DBLP:journals/dss/CamargoDG20}, a tool to discover a BPS model from an event log and, using hyperparameter optimization, tune it in order to minimize a loss function and obtain an enhanced BPS model that reproduce more faithfully the process behavior.
Following with the aim of enhancing simulation models, Pourbafrani et al.~\cite{DBLP:conf/caise/PourbafraniA21} proposed to use coarse-grained BPS models to simulate processes at a higher abstraction level, claiming that it eases the discovery of models mimicking the real behavior, and the exploration of alternative scenarios.
Estrada-Torres et al. proposed in~\cite{DBLP:journals/dke/Estrada-TorresC21} a method to discover BPS models in the presence of multitasking, and enhanced with support to resource unavailability.
Finally, Meneghello et al.~\cite{DBLP:conf/rcis/MeneghelloFLAT22} presented a framework to evaluate the accuracy of a BPS model w.r.t.\ the original event log, and give insights to improve this accuracy.
Nevertheless, none of these techniques address the problem of enhancing the BPS model by considering extraneous activity delays.

Camargo et al.~\cite{DBLP:conf/caise/CamargoDR22} note that existing data-driven BPS discovery methods only take into account waiting times due to resource contention and resource unavailability, while a great portion of waiting time is, in reality, caused by extraneous factors~\cite{DBLP:conf/pakdd/AndrewsW17}.
Camargo et al. propose to address this shortcoming by modeling waiting times using using deep learning models. They compare their performance against data-driven BPS models.
Although the results show that deep learning BPS models are more accurate, they cannot be used to perform ''what-if`` analysis, as deep learning models work as a black box.
On the contrary, in this study, we aim to produce data-driven BPS models that can be used for ''what-if`` analysis.

Martin et al.~\cite{DBLP:journals/is/MartinPM21} note that waiting times in business processes may be caused by batch processing. The authors identify different types of batching-related waiting times and propose an approach to discover such waiting times in event logs. Lashkevich et al.~\cite{DBLP:conf/caise/LashkevichMCSD23} build upon this approach to propose a method to separate the waiting time between pairs of consecutive activity instances into waiting times attributable to resource contention, to resource unavailability, to batch processing, and to case prioritisation -- the remaining waiting time is attributed to extraneous factors -- but modeling these extraneous delays is not in the scope of~\cite{DBLP:conf/caise/LashkevichMCSD23}.

Another related body of work is that of queue mining~\cite{DBLP:journals/is/SenderovichWGM15}, which propose an approach to model delays preceding the start of an activity due to queueing effects. This latter approach takes into account the fact that delays may be dependent of various factors, such as the type of customer and the queuing policy. However, the approach focused on delays due to resource contention.


%% file: approach.tex
\section{Extraneous Activity Delays Modeling\label{sec:approach}}

\begin{figure*}[t]
    \centering
    \includegraphics[width=0.9\textwidth]{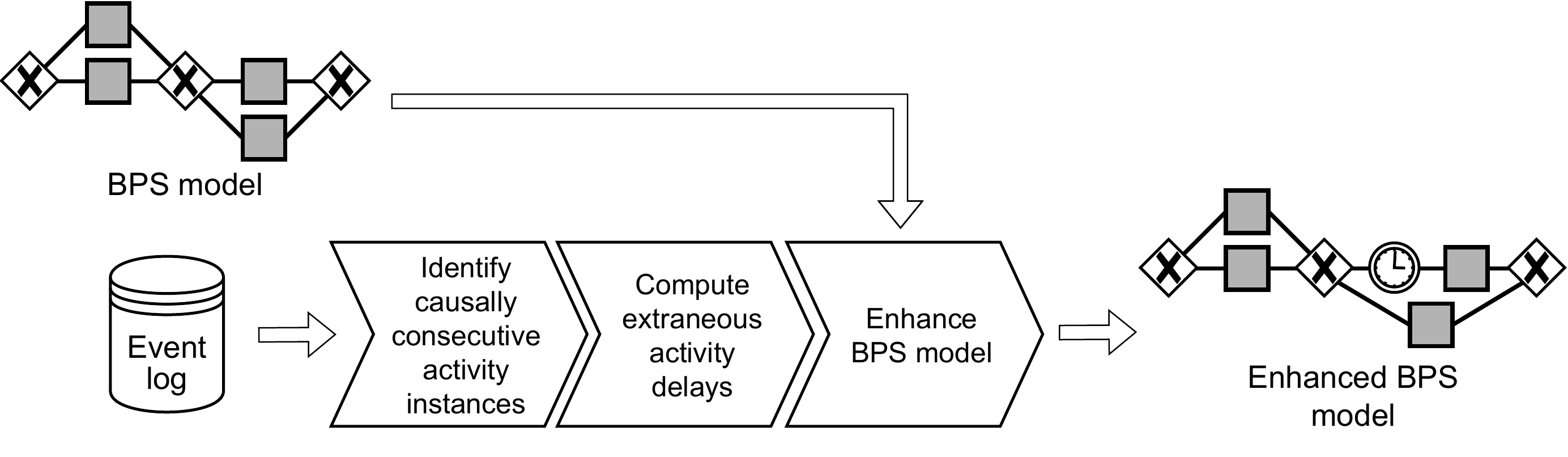}
    \caption{Overview of the approach presented in this paper.}
    \label{fig:proposal-overview}
\end{figure*}

This section describes the proposed approach to enhance a BPS model by adding timer events that capture the waiting times caused by extraneous factors.
\figurename~\ref{fig:proposal-overview} depicts the main structure of our proposal.
With the information recorded in the input event log, our approach first identifies the pairs of causally consecutive activity instances (\sectionname~\ref{subsec:causal-consec-identif}), i.e., for each activity instance $\varepsilon_{t}$ in the event log, the pair $(\varepsilon_{s}, \varepsilon_{t})$ such that $\varepsilon_{s}$ enabled $\varepsilon_{t}$.
Then, using the information about the availability of the resource that performed each $\varepsilon_{t}$, our approach estimates the amount of their waiting time associated to extraneous factors (\sectionname~\ref{subsec:extraneous-activity-delays}).
For this part of the approach, we propose two methods: \textit{i)} a technique that estimates the extraneous delay based in the last instant in which the resource was available to process $\varepsilon_{t}$ (\sectionname~\ref{subsubsec:naive}), and \textit{ii)} another technique that works under a more sophisticated reasoning to reduce the impact of resource unavailability periods overlapping with the extraneous delays (\sectionname~\ref{subsubsec:eclipse-aware}).
Finally, for each activity with discovered extraneous delays, a timer event is added to the BPS model received as input, with the duration distribution result of the observed extraneous delays (see \sectionname~\ref{subsec:model-enhancement}).

\subsection{Causally consecutive activity instance identification\label{subsec:causal-consec-identif}}

The first step to compute the waiting time associated to extraneous factors of an activity instance is to identify its total waiting time, which corresponds to the period of time since it became available to be processed (i.e., enabled), until it started to be processed (i.e., its start time).
In this context, an activity instance is not considered to be enabled until the instances of its causally preceding activities finish.
For example, in the event log from \tablename~\ref{tab:activity-instance-log-example}, an invoice is first registered, then posted and notified as accepted (both in parallel), and after that, paid.
In this case, the payment of the invoice cannot be made until it is both posted and notified as accepted.
Thus, activity \textit{``Pay invoice''} is not enabled for processing until both activities (\textit{``Post invoice''} and \textit{``Notify acceptance''}) finish.
Accordingly, we identify the pairs of all causally consecutive activity instances $(\varepsilon_{s}, \varepsilon_{t})$ from the event log.
In this context, $\varepsilon_{t}$ denotes the target activity, i.e., the activity that has been (potentially) delayed; and $\varepsilon_{s}$ denotes its causal predecessor, or source activity instance, i.e., the activity instance that enabled $\varepsilon_{t}$.
Then, the total waiting time associated with each pair $(\varepsilon_{s}, \varepsilon_{t})$ corresponds to the interval $\langle\tau_e(\varepsilon_{s}), \tau_s(\varepsilon_{t})\rangle$, i.e., the period of time since the end of $\varepsilon_{s}$ until the start of $\varepsilon_{t}$.

\begin{figure}[t]
    \centering
    \includegraphics[width=0.99\columnwidth]{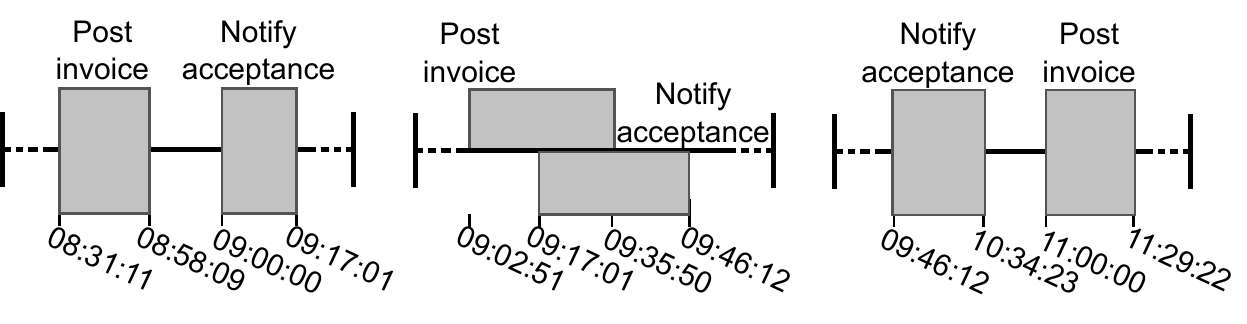}
    \caption{Example of concurrent activities executed in parallel and sequential order, corresponding to the traces (from left to right) 512, 513, and 514 in \tablename~\ref{tab:activity-instance-log-example}.}
    \label{fig:concurrency-example}
\end{figure}

In a sequential process, the causal predecessor of each activity instance is its chronologically preceding activity instance.
However, this observation does not hold if there are concurrency relations between activities in a process (a common situation in real-life processes).
A concurrency relation between two activities $A$ and $B$ ($A \parallel B$) denotes that these activities can overlap -- i.e., there is a period in time where both $A$ and $B$ are being executed at the same time --, or occur in any order -- i.e., sometimes $A$ following $B$ and other times $B$ following $A$ --, thus there is no causal relation between them.\footnote{We note that other notions of concurrency have been proposed in the field of process mining.
An in-depth treatment of concurrency notions in process mining is provided by Armas-Cervantes et al. in~\cite{DBLP:journals/toit/Armas-Cervantes19}. In this paper, we adopt the notion of overlapping concurrency between pairs of activities as defined above. The proposed approach is, however, modular and could be adapted to exploit other notions of concurrency.}
For example, in \tablename~\ref{tab:activity-instance-log-example}, activities \textit{``Post invoice''} and \textit{``Notify acceptance''} were recorded in parallel (trace 513) or sequentially in both orders (traces 512 and 514), as no one is required to start the other (see \figurename~\ref{fig:concurrency-example}).

Hence, to identify the causal predecessor of an activity instance, we first detect the concurrency relations between the activities of the process.
To do this, we develop on the idea proposed by Augusto et al. in~\cite{DBLP:conf/icpm/AugustoDR20}, and consider two activities to be concurrent when the proportion of times they overlap in time w.r.t.\ the number of times they co-occur is higher or equal to a given threshold $\zeta$.
In this way, once the concurrency relation between the activities is known, the causal predecessor of an activity instance is the closest previous activity instance not overlapping and not being concurrent to the current one.

\begin{defn}[Causal predecessor\label{def:causal-predecessor}]

    Given an activity instance log $\mathcal{L}$, and a concurrency oracle that states if two activities $\alpha_{i}, \alpha_{j} \in \mathcal{L}$ have ($\alpha_{i} \parallel \alpha_{j}$) or do not have ($\alpha_{i} \nparallel \alpha_{j}$) a concurrent relation, the set of \textit{predecessors} of an activity instance $\varepsilon = (\varphi, \alpha, \tau_{s}, \tau_{e}, \rho)$, such that $\varepsilon \in \mathcal{L}$, is defined as
    \begin{equation}\label{eq:predecessors}
    \begin{split}
        \mathcal{E}_{pre}(\varepsilon) \coloneqq \{& \varepsilon_{i} \in \mathcal{L} \ \mid \ \varphi(\varepsilon_{i}) = \varphi(\varepsilon) \ \wedge\ \\
        & \tau_{e}(\varepsilon_{i}) \leq \tau_{s}(\varepsilon) \ \wedge \ \alpha(\varepsilon_{i}) \nparallel \alpha(\varepsilon)\}
    \end{split}
    \end{equation}
    , i.e., the activity instances in $\mathcal{L}$, belonging to the same process trace, with an end time previous or similar to $\varepsilon$'s start time (i.e., preceding and not overlapping), and which activity does not have a concurrent relation with $\varepsilon$'s activity. Accordingly, the \textit{causal predecessor} of $\varepsilon$ is defined as
    \begin{equation}\label{eq:causal-predecessor}
    \begin{split}
        \varepsilon_{cp}(\varepsilon) \coloneqq
        \underset {
                \varepsilon_i\ \in\ \mathcal{E}_{pre}(\varepsilon)
        }{
            \operatorname {arg\,max}
        }\
        \tau_{e}(\varepsilon_{i})
    \end{split}
    \end{equation}
    , i.e., the predecessor of $\varepsilon$ with the maximum end timestamp.

\end{defn}

\begin{figure}[t]
    \centering
    \includegraphics[width=0.98\columnwidth]{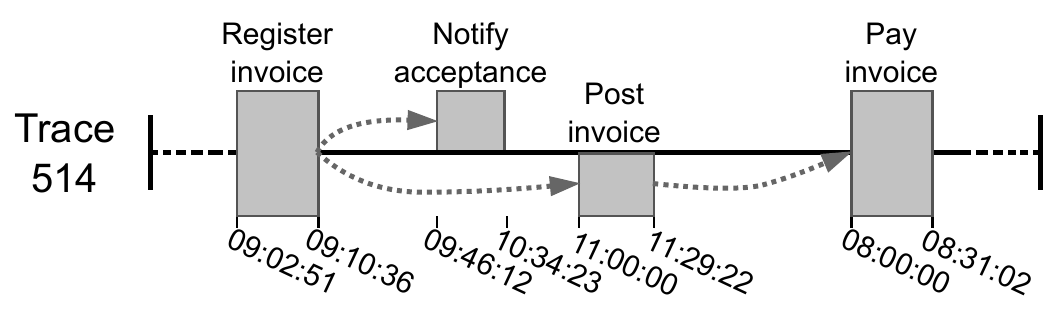}
    \caption{Timeline of trace 514 of the example in \tablename~\ref{tab:activity-instance-log-example}, where each gray box represents each activity instance, and each dotted arrow connects an activity instance (target) with its causal predecessor (source).}
    \label{fig:enablement-time}
\end{figure}

Then, all pairs of causally consecutive activity instances $\mathcal{E}_{CC} = \{(\varepsilon_{s,1}, \varepsilon_{t,1}),\ (\varepsilon_{s,2}, \varepsilon_{t,2}),\ \ldots,\ (\varepsilon_{s,n}, \varepsilon_{t,n})\}$ from a log $\mathcal{L}$ can be computed by applying Eq.~\ref{eq:causal-predecessor} for each activity instance $\varepsilon_{t,i} \in \mathcal{L}$, such that $ \varepsilon_{s,i} = \varepsilon_{cp}(\varepsilon_{t,i})$.
As an example, \figurename~\ref{fig:enablement-time} shows the timeline of the process trace 514 from the log in \tablename~\ref{tab:activity-instance-log-example}, depicting the pairs of causally consecutive activity instances.
As it can be seen, both \textit{``Post invoice''} and \textit{``Notify acceptance''}, which have a concurrency relation between them, have \textit{``Register invoice''} as causal predecessor.
Finally, the causal predecessor of \textit{``Pay invoice''} is \textit{``Post invoice''}, the previous non-overlapping activity instance not being concurrent to it.

It must be noted that we do not compute the pairs of causally consecutive activity instances for the $\varepsilon_{t, i}$ that do not have a causal predecessor.
These cases correspond to two types of activity instances: \textit{i)} the first activity instance of each trace, and \textit{ii)} the activity instances with all predecessors being concurrent or overlapping them.
In both cases, the causal predecessor of $\varepsilon_{t, i}$ is the trace arrival.
Martin et al.~\cite{DBLP:conf/bpm/MartinDC15} and Berkenstadt et al.~\cite{DBLP:conf/icpm/BerkenstadtGSSW20} have presented different techniques to estimate the arrival times of process traces that could be used as enablement time for this activity instances.
Nevertheless, this is out of the scope of this paper.

\subsection{Extraneous Activity Delays Computation\label{subsec:extraneous-activity-delays}}

Once the set of pairs of causally consecutive activity instances $\mathcal{E}_{CC} = \{(\varepsilon_{s,1}, \varepsilon_{t,1}),\ (\varepsilon_{s,2}, \varepsilon_{t,2}),\ \ldots,\ (\varepsilon_{s,n}, \varepsilon_{t,n})\}$ is known, the waiting time associated with each of them can be easily computed as the interval from $\varepsilon_{s,i}$'s end time to $\varepsilon_{t,i}$'s start time, for $i \in 1\ldots n$.
Nevertheless, not all this waiting time is caused by extraneous factors.
For example, if no resources are available, an enabled activity instance must wait until a resource can be assigned to them.
When the resources are working on other activities, this waiting time is denoted as resource contention.
If the resources are out of their working schedules, it is denoted as resource unavailability.
However, both resource contention and unavailability waiting times are already modeled during the simulation by the resource allocation model and, thus, they must not be considered as extraneous delays.
In this section, we propose two methods to compute the waiting time associated with extraneous factors.

\begin{figure*}[t]
    \centering
    \includegraphics[width=0.98\textwidth]{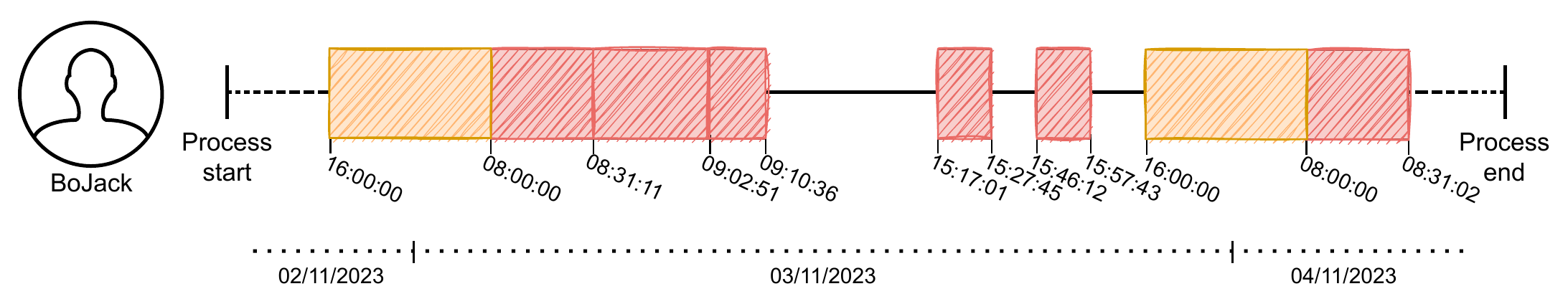}
    \caption{Individual timeline of the resource ``BoJack'' corresponding to the example in \tablename~\ref{tab:activity-instance-log-example}, where the empty intervals represent available periods, and the orange and red striped boxes represent, respectively, non-working periods and other activity instances performed by this resource.}
    \label{fig:resource-availability}
\end{figure*}

\subsubsection{Naive method\label{subsubsec:naive}}

\vspace{5pt}

Our first proposal is based on the reasoning that, as soon as an activity instance is enabled, and a resource (who can perform it) is available, she/he/they will start working in the activity instance.
A similar reasoning is also used by Zabka et al. in \cite{ZabkaBA2021}, where they consider the end of each activity instance as the start of the next one performed by the same resource. 
However, they only consider resources not being available when they are performing other activities, not taking into account their working schedules -- e.g., scheduled breaks.
We develop on this idea and propose to consider both causes of resource unavailability by building a timeline for each resource with their non-working periods, and the recorded activity instances that they performed.
\figurename~\ref{fig:resource-availability} shows the timeline of a resource of the activity instance log depicted in \tablename~\ref{tab:activity-instance-log-example}.

We use the resource calendar information present in the BPS model to obtain the non-working periods of each resource.
Typically, BPS models report the working periods of the resources in a weekly manner -- e.g., the working schedules each day of the week.
For each resource, we compute the non-working (absolute) intervals $NW(\rho) = \{ \langle\tau_{i,1}, \tau_{j,1}\rangle,\ \langle\tau_{i,2}, \tau_{j,2}\rangle,\ \ldots,\ \langle\tau_{i,m}, \tau_{j,m}\rangle\}$ as all the intervals contained in the timeline of the process, i.e., $ \forall_{\langle\tau_i, \tau_j\rangle \in NW(\rho)} : \tau_i \geq min(\{\tau_{s}(\varepsilon_{i}) \mid \varepsilon_{i} \in \mathcal{L}\}) \wedge \tau_j \leq max(\{\tau_{e}(\varepsilon_{i}) \mid \varepsilon_{i} \in \mathcal{L}\}) $, that do not overlap with their working schedules.
If the resources' working schedules are not present in the BPS model, they can be estimated with discovery techniques such as the one presented in~\cite{DBLP:conf/bpm/Lopez-PintadoD22}, which analyzes the instants in time when each resource interacted with the system -- i.e., the start and end of each activity instance -- to build a weekly working calendar composed of time intervals in which there was enough evidence, based on given support and confidence values, that the resource is working.

In this way, the \textit{resource availability time} of an activity instance corresponds to the latest of both \textit{i)} the end of the previous non-working period of its resource, or \textit{ii)} the end time of the previous activity instance performed by its resource.
Accordingly, the resource availability time ($\tau_{rat}$) of an activity instance is defined as follows:

\begin{defn}[Resource Availability Time\label{def:resource-availability-time}]

    Given an activity instance log $\mathcal{L}$, the \textit{resource availability time} of an activity instance $\varepsilon = (\varphi, \alpha, \tau_{s}, \tau_{e}, \rho)$, such that $\varepsilon \in \mathcal{L}$, is defined as
    \begin{equation}
    \begin{split}
        \tau_{rat}(\varepsilon) \coloneqq \ &
        max(\{ \
            \tau_{e}(\varepsilon_{i}) \ \mid \
            \varepsilon_{i} \in \mathcal{L} \ \wedge \
            \rho(\varepsilon_{i}) = \rho(\varepsilon) \ \wedge \\
            & \tau_{e}(\varepsilon_{i}) \leq \tau_{s}(\varepsilon)
        \ \} \ \cup \ \{ \
            \tau_j \ \mid \
            \tau_j \leq \tau_s(\varepsilon) \ \wedge \\
             & \langle\tau_i, \tau_j\rangle \in NW(\rho(\varepsilon))
        \ \})
    \end{split}
    \end{equation}
    , i.e., the largest of both \textit{i)} the end times of the activity instances of $\mathcal{L}$, processed by the same resource as $\varepsilon$, being previous or equal to its start (not overlapping); and \textit{ii)} the end of the resource's non-working periods (the time instants where they became available for processing) that are smaller or equal to $\varepsilon$'s start.

\end{defn}


In order to discard the waiting times due to resource contention and resource unavailability from the extraneous waiting time calculation, we propose to consider the resource availability time to compute the earliest instant at which an activity instance could have started.
To do this, we define the earliest possible start time of an activity instance (henceforth, earliest start time for short) as the first instant in time when both its causal predecessor finished, and its resource became available to process it.
Accordingly, the waiting time caused by extraneous factors can be easily computed as the difference between their recorded start time and their earliest start time.

\begin{defn}[Naive extraneous activity delay\label{def:naive-extraneous-delay}]

    Given an activity instance log $\mathcal{L}$ of a process with a set of activities $A$, the \textit{naive extraneous delay} associated with a causally consecutive pair of activity instances $(\varepsilon_{s}, \varepsilon_{t})$, such that $\varepsilon_{s}, \varepsilon_{t} \in \mathcal{L}$, is defined as
    \begin{equation}\label{eq:naive-extraneous-delay}
        \begin{split}
            \omega^{\prime}_{ex}((\varepsilon_{s}, \varepsilon_{t})) \coloneqq \tau_{s}(\varepsilon_{t}) - max(\tau_{rat}(\varepsilon_{t}), \tau_{e}(\varepsilon_{s}))
        \end{split}
    \end{equation}
    , i.e., the period of time from its earliest start time to its recorded start.

\end{defn}

\begin{figure}[t]
    \centering
    \includegraphics[width=0.98\columnwidth]{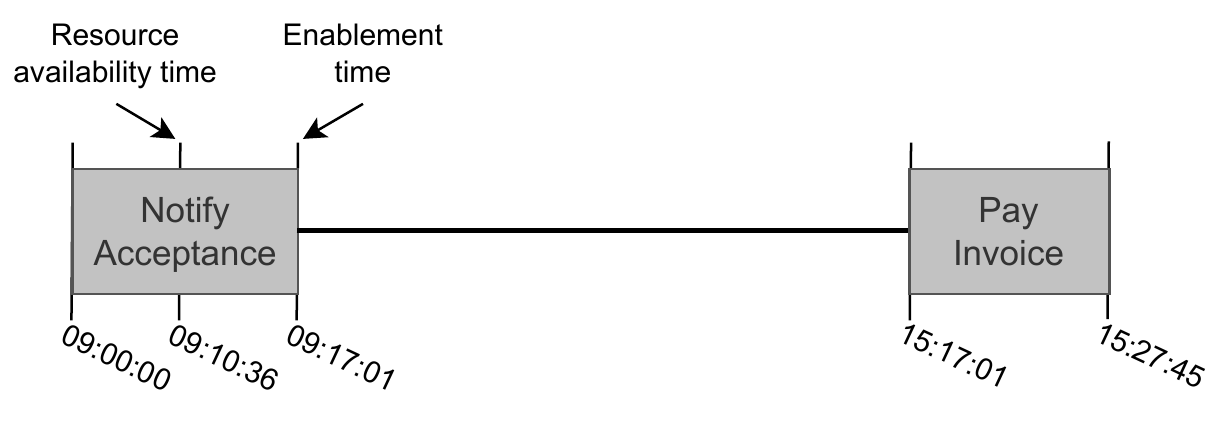}

    \includegraphics[width=0.98\columnwidth]{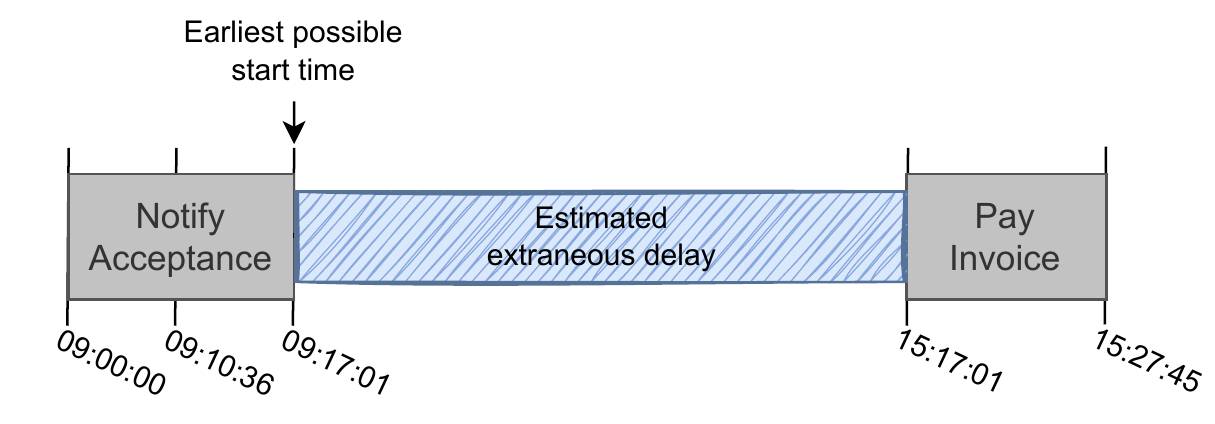}

    \caption{Timeline denoting the resource availability, enablement, and earliest start times for the pair of causally consecutive activity instances of the activities ``Notify acceptance'' and ``Pay invoice'' of the process trace 512, corresponding to the example in \tablename~\ref{tab:activity-instance-log-example}.
    The blue striped box denotes the estimated waiting time due to extraneous factors with the naive technique.}
    \label{fig:naive-extraneous-delay}
\end{figure}

\figurename~\ref{fig:naive-extraneous-delay} depicts an example of the estimation of the extraneous delays for one pair of causally consecutive activity instances of the example in \tablename~\ref{tab:activity-instance-log-example}.
In this example, the resource was already available when the source activity instance completed and, thus, the computed extraneous delay corresponds to the total waiting time (there is no resource unavailability or resource contention involved).

\subsubsection{Eclipse-aware method\label{subsubsec:eclipse-aware}}

\vspace{5pt}

The previous proposal relies on the reasoning that, from the earliest start of an activity instance, its resource is available to process the activity, and the activity is ready to be processed.
Thus, if the activity is not being processed, it must be due to an unknown cause of waiting time -- i.e., extraneous factors.
Although this reasoning holds in this direction, it might not be the case in the opposite -- i.e., the resource might not be available during all extraneous waiting time. 
For example, while an activity is waiting for, e.g., the customer to provide additional documents, the resource may perform other activity instances, or even finish her/his/their working hours of that day.
When this happens, the waiting time caused by resource contention or unavailability \textit{eclipses}, either partially or totally, the waiting time due to extraneous factors.
Two types of interaction can happen in these cases, \textit{i)} there is extraneous waiting time after the resource finishes the previous activity, so the discovered delay has a shorter duration than the real one (\textit{partial eclipse}), and \textit{ii)} the extraneous waiting time ends earlier than the resource contention, so the activity starts as soon as its resource is available to process it, and no extraneous delay is discovered (\textit{total eclipse}).

\begin{figure*}[t]
    \centering
    \includegraphics[width=0.48\textwidth]{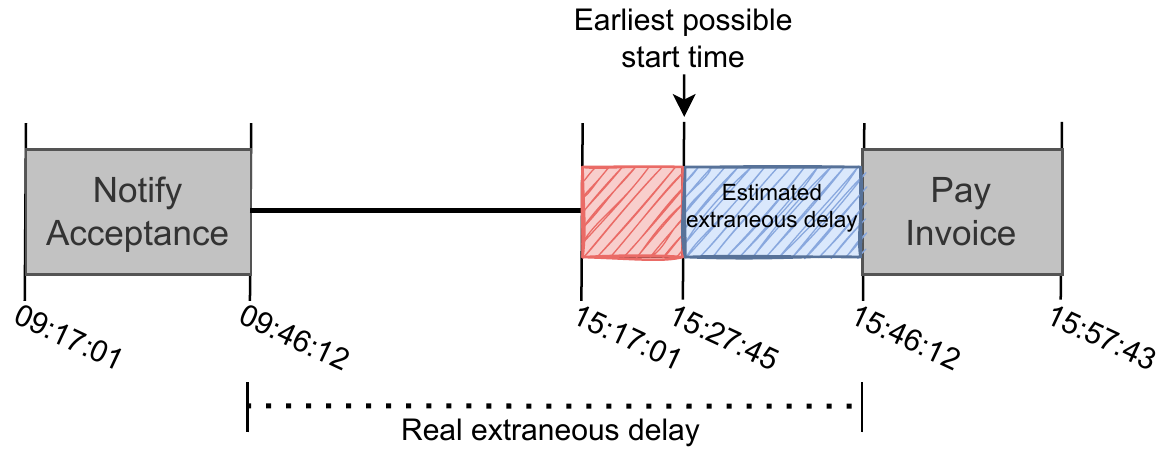}
    \hfill
    \includegraphics[width=0.48\textwidth]{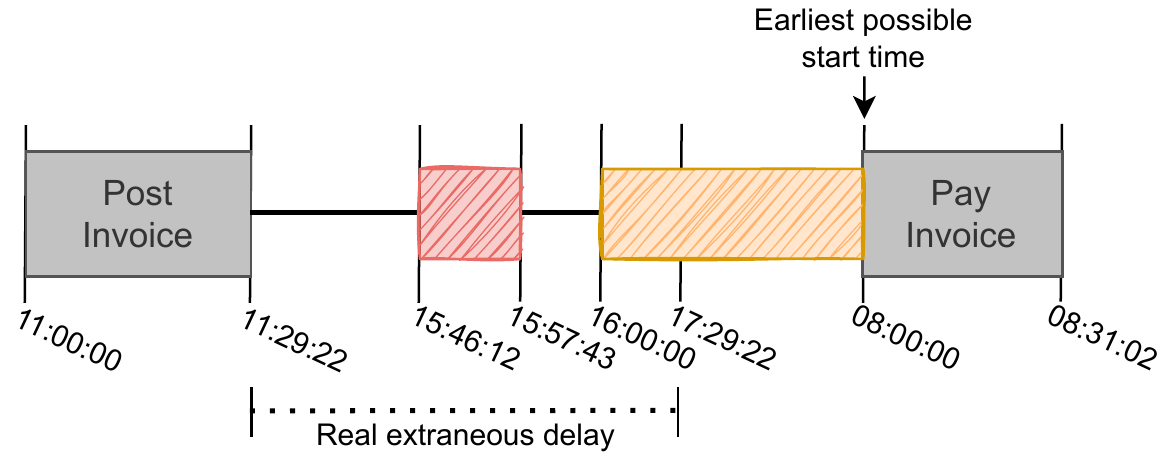}

    \caption{Example of a partial eclipse (left) and a total eclipse (right) observed for activity instance ``Pay Invoice'' of traces 513 and 514 of \tablename~\ref{tab:activity-instance-log-example}.
    The blue striped box represents the extraneous delay estimated by the Naive proposal (0 in the total eclipse).
    The red and orange striped boxes represent, respectively, other activity instances performed by this resource and non-working periods.}
    \label{fig:example-eclipses}
\end{figure*}

\figurename~\ref{fig:example-eclipses} depicts two eclipse examples for two causally consecutive pairs of activity instances.
In trace 513 (left), the execution of another activity instance (``Pay invoice'' from trace 512, henceforth $\varepsilon_{o}$) partially eclipses the extraneous waiting time of the target activity instance (``Pay invoice'', henceforth $\varepsilon_{t}$).
While $\varepsilon_{t}$ was waiting due to extraneous factors, its resource performed $\varepsilon_{o}$.
For this reason, $\varepsilon_{t}$'s earliest start is set to the end time of $\varepsilon_{o}$, causing the naive technique to underestimate the extraneous delay.
In the pair of activity instances of trace 514 (right), the extraneous delay finished while the resource that performed $\varepsilon_{t}$ (``Pay invoice'') was off-duty.
Thus, $\varepsilon_{t}$ waited until 04/11/2023 08:00:00 for the resource to be available again.
This latter example shows a total eclipse, where the earliest start time is similar to $\varepsilon_{t}$'s start and, thus, the naive technique reasoning shows no extraneous delay.

In order to reduce the impact of eclipses in the extraneous delay estimation, we propose a second method (Eclipse-aware), that works on a more sophisticated reasoning.
For each pair of causally consecutive activity instances $(\varepsilon_{s}, \varepsilon_{t})$, we compute the first and last instants in time in which $\varepsilon_{t}$ could have started -- i.e., $\varepsilon_{s}$ completed, and its resource was available to process it.
Following the same reasoning -- i.e., if $\varepsilon_{t}$ did not start is due to extraneous factors --, these two instants denote the first and last instants with evidence that $\varepsilon_{t}$ was waiting due to extraneous factors.
Thus, we conclude that the extraneous delays of $\varepsilon_{t}$ comprise, at least, the period between these two instants.


\begin{figure*}[t]
    \centering
    \includegraphics[width=0.48\textwidth]{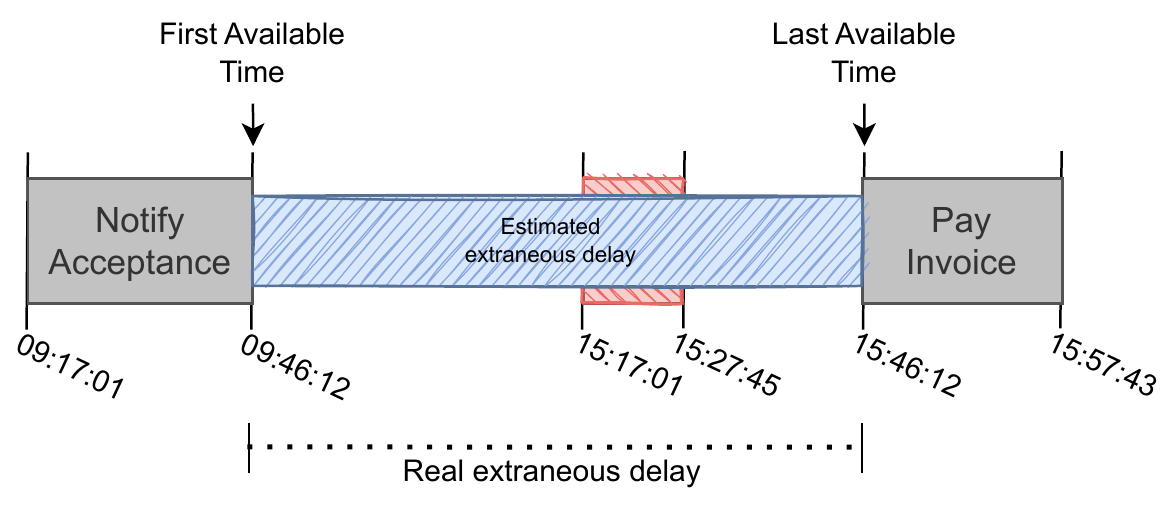}
    \hfill
    \includegraphics[width=0.48\textwidth]{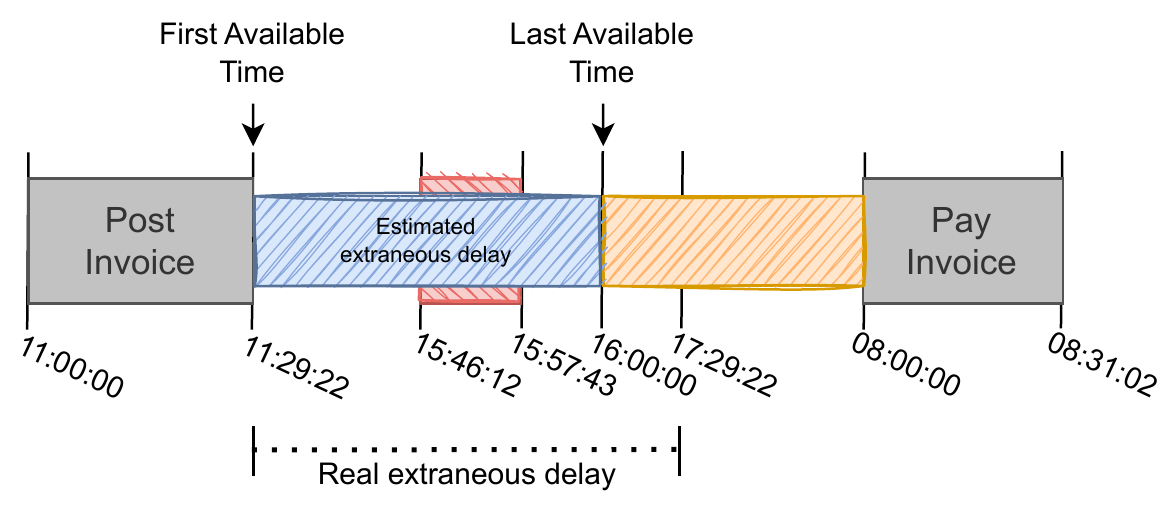}

    \caption{Timeline denoting the first and last available times, the start and end times, and the real extraneous delay for the activity instance ``Pay invoice'' of traces 513 and 514 of \tablename~\ref{tab:activity-instance-log-example}.
    The blue striped box represents the extraneous delay estimated by the Eclipse-aware proposal.
    The red and orange striped boxes represent, respectively, other activity instances performed by this resource and non-working periods.}
    \label{fig:complex-extraneous-delay}
\end{figure*}

\begin{defn}[First and Last Available Times\label{def:first-last-available-time}]

    Given an activity instance log $\mathcal{L}$, a period of time $\lambda$, and a pair of causally consecutive activity instances $(\varepsilon_{s}, \varepsilon_{t})$, such that $\varepsilon_{s}, \varepsilon_{t} \in \mathcal{L}$, we define the availability intervals associated to it as follows
    \begin{equation}
        \begin{split}
            AI((\varepsilon_{s}, \varepsilon_{t})) \coloneqq \ & \{ \
                \langle\tau_i, \tau_j\rangle \mid
                \tau_j - \tau_i \geq \lambda \ \wedge \\
                & \langle\tau_i, \tau_j\rangle \models \langle\tau_{e}(\varepsilon_{s}), \tau_s(\varepsilon_{t})\rangle \ \wedge \\
                & (\nexists_{\langle\tau_k, \tau_l\rangle \in NW(\rho(\varepsilon_{t}))} : \langle\tau_i, \tau_j\rangle \perp \langle\tau_k, \tau_l\rangle) \ \wedge \\
                & (\nexists_{\varepsilon_i \in \mathcal{L}} : \rho(\varepsilon_i) = \rho(\varepsilon) \ \wedge \\
                & \langle\tau_i, \tau_j\rangle \perp \langle\tau_s(\varepsilon_i), \tau_e(\varepsilon_i)\rangle)
            \ \}
        \end{split}
    \end{equation}
    , i.e., the set of intervals of duration greater or equal to $\lambda$, contained in the waiting time of $(\varepsilon_{s}, \varepsilon_{t})$, and such that the resource that performed $\varepsilon_{t}$ is neither out of its working periods, nor processing any other activity instance.
    Consequently, the \textit{first available time} and \textit{last available time} are defined as
    \begin{equation}
        \tau_{fat}((\varepsilon_{s}, \varepsilon_{t})) \coloneqq min(\{ \ \tau_i \mid \langle\tau_i, \tau_j\rangle \in AI((\varepsilon_{s}, \varepsilon_{t})) \ \})
    \end{equation}
    \begin{equation}
        \tau_{lat}((\varepsilon_{s}, \varepsilon_{t})) \coloneqq max(\{\  \tau_j \mid \langle\tau_i, \tau_j\rangle \in AI((\varepsilon_{s}, \varepsilon_{t})) \ \})
    \end{equation}
    , i.e., the start of the first and the end of the last of these intervals, respectively.

\end{defn}

We consider the resource to be available only when the periods exceed a minimum duration ($\lambda$) to avoid the noise of short breaks.
For example, if the resource is available from 8:00 AM, but logs the start of the first activity in the system at 8:03 am; or when she/he/they registers the end of an activity at 11:00 AM, and logs the start of the next one at 11:04 AM.
In this way, by setting $\lambda$ to 5 minutes, we do not consider available periods of less than 5 minutes between one unavailable interval and the next.

Once the first and last available times associated with a pair of causally consecutive activity instances are known, we estimate its extraneous delay as the interval between those two instants.
Accordingly, the extraneous activity delay of an activity instance is defined as follows:

\begin{defn}[Eclipse-aware extraneous activity delay\label{def:complex-extraneous-delay}]

    Given an activity instance log $\mathcal{L}$ of a process with a set of activities $A$, the \textit{eclipse-aware extraneous delay} of a pair of causally consecutive activity instances $(\varepsilon_{s}, \varepsilon_{t})$, such that $\varepsilon_{s}, \varepsilon_{t} \in \mathcal{L}$, is defined as
    \begin{equation}
        \omega^{\prime\prime}_{ex}((\varepsilon_{s}, \varepsilon_{t})) = \tau_{lat}((\varepsilon_{s}, \varepsilon_{t})) - \tau_{fat}((\varepsilon_{s}, \varepsilon_{t}))
    \end{equation}
    , i.e., the time period from the first available time until the last available time of $\varepsilon_{t}$.

\end{defn}

\figurename~\ref{fig:complex-extraneous-delay} depicts two examples of this computation for traces 513 and 514 of the running example from \figurename~\ref{tab:activity-instance-log-example}.
In trace 513, the partial eclipse consists of the same resource executing another activity instance during the extraneous waiting time of the current one.
This would cause an underestimation by the naive technique, as the earliest start time would be estimated at the end of this activity.
However, the estimated extraneous delay is similar to the real one by using the first and last available instants.
In trace 514, the total eclipse hides the instant when the extraneous waiting time ended.
In this case, the resource was unavailable from 03/11/2021 at 15:46:12 to 04/11/2021 at 08:00:00, and the extraneous delay ended on 03/11/2021 at 17:29:22.\footnote{The use of a minimum duration ($\lambda$) prevents the technique from wrongly considering a short break from one working item and the next one (e.g., from the end of an activity at 15:57:43, to the start of a non-working period at 16:00:00) as an available period where she/he/they would start the activity under analysis if enabled.}
As there is no evidence of the resource being available to start the activity from 03/11/2021 at 15:46:12 (the last available time), our proposal is to approximate the delay by using this instant.

\subsubsection{Extrapolated eclipse-aware method\label{subsubsec:eclipse-aware-extrapolation}}

\vspace{5pt}

\begin{figure}[t]
    \centering
    \includegraphics[width=0.98\columnwidth]{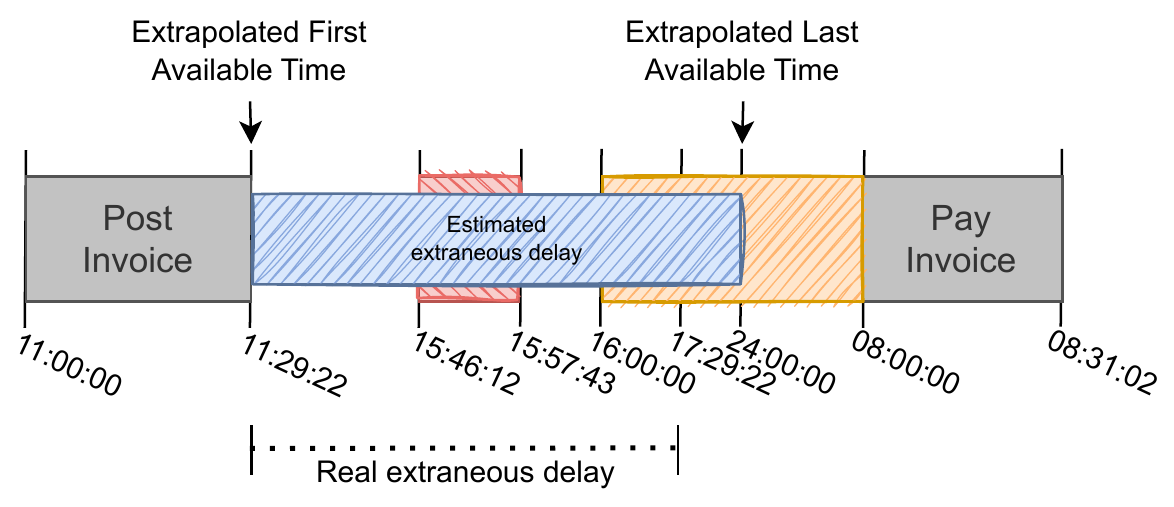}

    \caption{Extrapolated first and last available times corresponding to the total eclipse example (right) in \figurename~\ref{fig:complex-extraneous-delay}.
    The blue, red, and orange striped boxes represent, respectively, \textit{i)} the extraneous delay estimated by the Extrapolated eclipse-aware proposal, \textit{ii)} other activity instances performed by the same resource that performed ``Pay Invoice'', and \textit{iii)} her/his/their non-working periods.}
    \label{fig:extrapolated-complex-extraneous-delay}
\end{figure}

As described in the previous subsection, the first and last available times denote, respectively, the first and last instants in which there is evidence that the resource could have started working in the activity, but did not.
However, this does not ensure that the extraneous delay started/ended at those instants.
For example, in some cases, the end of the extraneous delay corresponds to an instant between the last available time, and the start of the activity instance (see \figurename~\ref{fig:extrapolated-complex-extraneous-delay}).
In these situations, estimating the end of the extraneous delay as the last available time (16:00:00) results in an average error of 1/2 the interval between this instant (16:00:00) and the start of the activity instance (08:00:00).
In order to reduce the average error of the estimation, we propose a third method that extrapolates the first and last available times to the middle point between them and the enablement and start, respectively, of the (potentially) delayed activity instance.
Then, the extrapolated eclipse-aware extraneous delay is defined as the difference between these two adjusted instants.

\begin{defn}[Extrapolated eclipse-aware extraneous activity delay\label{def:complex-extrapolated-extraneous-delay}]

    Given an activity instance log $\mathcal{L}$ of a process with a set of activities $A$, and a pair of causally consecutive activity instances $(\varepsilon_{s}, \varepsilon_{t})$, such that $\varepsilon_{s}, \varepsilon_{t} \in \mathcal{L}$; its \textit{extrapolated first available time} and \textit{extrapolated last available time} are defined as
    \begin{equation}
        \tau_{fat}^{\prime}((\varepsilon_{s}, \varepsilon_{t})) \coloneqq \tau_{fat}((\varepsilon_{s}, \varepsilon_{t})) - (\tau_{fat}((\varepsilon_{s}, \varepsilon_{t})) - \tau_{e}(\varepsilon_{s}))\ /\ 2
    \end{equation}
    \begin{equation}
        \tau_{lat}^{\prime}((\varepsilon_{s}, \varepsilon_{t})) \coloneqq \tau_{lat}((\varepsilon_{s}, \varepsilon_{t})) + (\tau_{s}(\varepsilon_{t}) - \tau_{lat}((\varepsilon_{s}, \varepsilon_{t})))\ /\ 2
    \end{equation}
    , i.e., the instant in the middle between the end of $\varepsilon_{s}$ and the first available time of $\varepsilon_{t}$, and the instant in the middle between the last available time of $\varepsilon_{t}$ and its start.
    Accordingly, the \textit{extrapolated eclipse-aware extraneous delay} of $(\varepsilon_{s}, \varepsilon_{t})$ is defined as
    \begin{equation}
        \omega^{\prime\prime\prime}_{ex}((\varepsilon_{s}, \varepsilon_{t})) = \tau_{lat}^{\prime}((\varepsilon_{s}, \varepsilon_{t})) - \tau_{fat}^{\prime}((\varepsilon_{s}, \varepsilon_{t}))
    \end{equation}
    , i.e., the period from the extrapolated first available time until the extrapolated last available time of $\varepsilon_{t}$.

\end{defn}

\subsection{BPS model enhancement\label{subsec:model-enhancement}}

The last step of the proposed approach is to enhance the BPS model by adding (duration-based) timer events to model the extraneous activity delays.
To this end, we first group the discovered extraneous activity delays and fit a duration distribution to assign to each timer event.
We propose two alternatives when performing this grouping: \textit{i)} the first approach considers the fact that the extraneous activity delays occur after the completion of an activity (\textit{ex-post}), e.g., a worker contacting a customer after recording a purchase order change, and blocking the execution flow until the customer is contacted; and \textit{ii)} the second approach considers the extraneous delays to occur before the start of an activity (\textit{ex-ante}), e.g., a worker waits for oral confirmation of the shipment address prior to triggering a shipment.

Accordingly, given the set of pairs of causally consecutive activity instances $\mathcal{E}_{CC} = \{(\varepsilon_{s}, \varepsilon_{t})\}$ recorded in the event log, we compute the multiset of ex-ante extraneous activity delays of each activity $\alpha_{i} \in A$ as
\begin{equation}\label{eq:set-extraneous-delays}
\begin{split}
    \Omega_{ex}(\alpha_{i})\ = \ \{\!\{\ & \omega_{ex}((\varepsilon_{s}, \varepsilon_{t})) \ \mid \
        (\varepsilon_{s}, \varepsilon_{t}) \ \in \ \mathcal{E}_{CC} \ \wedge \ \\
        & \alpha(\varepsilon_{t}) = \alpha_{i} \ \}\!\}
\end{split}
\end{equation}
by using $\omega^{\prime}_{ex}$ for the naive approach, and $\omega^{\prime\prime}_{ex}$ for the eclipse-aware proposal.
Similarly, the set of ex-post extraneous activity delays can be computed by applying Eq.~\ref{eq:set-extraneous-delays} changing the last condition to $\alpha(\varepsilon_{s}) = \alpha_{i}$.

In order to discard the presence of outliers, i.e., an activity with a scarce number of delays w.r.t.\ their total number of executions, we consider an outlier threshold ($\delta$) to inject a timer event only to those activities in which $\lvert\{ \omega_{ex} \mid \omega_{ex} \in \Omega_{ex}(\alpha) \wedge \omega_{ex} > 0 \}\rvert\ /\ \lvert \Omega_{ex}(\alpha) \rvert > \delta$, i.e., the proportion of positive extraneous delays discovered is higher than $\delta$.


Finally, to correct potential deviations due to the presence of eclipses, we propose an additional step that considers the presented approaches as initial estimations, and introduces a scale factor $\gamma_{i}$ for each activity $\alpha_{i}$, such that all its discovered extraneous delays are scaled as $\Omega^{\prime}_{ex}(\alpha_{i}) = \{\!\{\gamma_{i} \cdot \omega_{ex} \mid \omega_{ex} \in \Omega_{ex}(\alpha_{i}) \}\!\}$.
Then, given a multiset $\Gamma = \{\!\{\gamma_{0}, \gamma_{1}, ..., \gamma_{n}\}\!\}$, corresponding to the set of activities $\{\alpha_{0}, \alpha_{1}, ..., \alpha_{n}\} \in A$ with discovered extraneous delays, we launch a Tree-structured Parzen Estimator~\cite{DBLP:conf/nips/BergstraBBK11} (TPE) hyperparameter optimization process to obtain the multiset $\Gamma$ that leads to a better simulation.
To avoid the possibility of overfitting, we split the input event log into two partitions, training and validation sets, discovering the extraneous delays from the former, and evaluating each solution in the latter.
To perform this split, we order the activity instances by end time and take the first 50\% for the train set, and the other 50\% for the validation set.\footnote{We perform a strict split considering partial traces to minimize the impact in the training-validation process. If we exclude complete traces, some resources might appear to be available in the training set, when actually they're busy working on a previous activity of a trace that is in the validation set, and that would affect the calculation of the resource availability times.}
Before the start of the optimization process, the $\Omega_{ex}$ are estimated using the training set.
Then, in each iteration of the optimization, the $\Omega^{\prime}_{ex}$ of each activity are computed based on the multiset $\Gamma$ given by the hyperparameter optimization algorithm, and the original BPS model is enhanced with the scaled extraneous delays.
We evaluate each iteration by comparing the enhanced BPS model with the validation set, using the Relative Event Distribution distance (described in \sectionname~\ref{subsec:rq3}) as distance measure to minimize by the TPE optimization.
In this way, the TPE hyperparameter optimizer returns the multiset $\Gamma$ values that lead to the enhanced BPS model that more closely reflects the temporal behavior of the log.

\subsection{Limitations}\label{subsec:limitation}

The techniques presented above rely on the idea of inserting a timer event prior to each activity in the process model, whenever we detect that the start times of the instances of this activity are affected by delays not attributable to resource contention or resource unavailability.

An underpinning assumption of this approach is that the extraneous delays are dependent on the activity type, for example, prior to performing an activity ``Close Claim'', a worker waits for at least 14 days for statutory reasons, or prior to starting an activity ``Review application'', a worker waits for the client to send some additional information (where the receipt of this information is not recorded explicitly in the event log).

As a side effect, the proposed approach also models extraneous waiting times caused by resource behavior (e.g., a resource going on unscheduled coffee breaks, or a resource being busy performing work in another process). Indeed, if resources take sporadic breaks, the approach will model the distribution of these waiting times on an aggregate basis, across all resources who perform a given activity.
Arguably, such resource-dependent waiting times, would be better modeled by attaching delays to the resources themselves, instead of attaching to the activities. This possible weakness of the approach is discussed in Section~\ref{sec:evaluation}.

A second limitation of the approach is that it assumes that resources do not engage in multitasking. Indeed, the proposed approach assumes that an extraneous delay occurs when an activity instance $\varepsilon_{i}$ is enabled, the resource who performs this activity is on-duty, and the resource is not performing another activity instance. If this resource engages in multitasking, then the fact that this resource is busy performing another activity instance is not a valid reason for not starting the $\varepsilon_{i}$. To capture multitasking, we would need to explicitly model the multitasking capacity of each resource (e.g., how many activity instances a resource may perform simultaneously). This is beyond the scope of this paper.

%% file: evaluation.tex
\section{Evaluation\label{sec:evaluation}}

This section reports on an experimental evaluation to analyze \textit{i)} the ability of the proposed technique to accurately discover already-known extraneous delays; \textit{ii)} the impact of modeling ex-post versus ex-ante delays; \textit{iii)} the extent to which BPS models, enhanced with different variants of the proposed approach, more closely replicate the temporal dynamics in a log, relative to a baseline BPS model without extraneous delays; and \textit{iv)} the computational efficiency of the different variants of our proposal.


With this purpose, four evaluation questions are presented.
For the first evaluation question, we focus on analyzing the accuracy of the ex-ante variants of our proposal when re-discovering extraneous activity delays in simulated scenarios in which they are known.\footnote{For simplicity, we focus on the ex-ante variant of our proposal in EQ1, and analyze the differences between the ex-post and ex-ante proposals in EQ2.}

\begin{quote}
    \textbf{Evaluation Question 1 (EQ1)}: \textit{Are the different variants of the proposed approach able to re-discover extraneous activity delays that are known to be present in the process from which the log was produced?}
\end{quote}

The purpose of this evaluation is to investigate the relative accuracy of the ex-ante variants of our proposal in the idealized scenario where the delay preceding each activity instance is dependent only on the activity type, and not on the resource who performs the activity instance. In other words, this experiment targets the scenario where the first limitation outlined in Section~\ref{subsec:limitation} does not apply.


For this evaluation question, we expect the eclipse-aware proposals to discover more accurate delays than the Naive proposal, as the latter one is more sensible to the impact of eclipses.
However, we expect that the three approaches re-discover all the extraneous activity delays with variations in their duration due to, as we have explained in \sectionname~\ref{subsec:model-enhancement}, eclipses in the waiting time.

As explained in \sectionname~\ref{subsec:model-enhancement}, we propose two alternatives when attributing the discovered extraneous delays to an activity instance (ex-post and ex-ante).
We foresee that the impact of this option when re-discovering already-known extraneous delays is rather conceptual -- in some scenarios, the extraneous delays occur after an activity, in others before -- than performance-related.
Thus, for the second evaluation question, we seek to analyze the impact that each option has on the discovery of extraneous delays.

\begin{quote}
\textbf{Evaluation Question 2 (EQ2)}: \textit{In processes with known activity delays, what is the relative performance of the ex-post and ex-ante variants of the proposed approach?}
\end{quote}

We expect two situations: \textit{i)} in the scenarios where the delay is located between two transitions, we expect both proposals to re-discover it as is; \textit{ii)} however, in the scenarios where the delays are located before or after a gateway, we expect one of the proposals to scatter the delay among the different gateway branches.

The third evaluation question focuses on the impact of the insertion of the discovered extraneous activity delays in real-life settings, where the first limitation outlined in Section~\ref{subsec:limitation} is likely to apply.

\begin{quote}
\textbf{Evaluation Question 3 (EQ3)}: \textit{In the context of real-life processes, given an event log and a BPS model of the process, do the proposed techniques improve the temporal accuracy of a BPS model?}
\end{quote}

Our proposal consists in enhancing the original BPS model by adding delays that are not modeled by resource contention or availability calendars.
Accordingly, we expect the enhancement of our proposal to improve the temporal accuracy w.r.t.\ the original BPS model.

Finally, the fourth evaluation question focuses on the analysis of the computational efficiency of the proposed approaches.
\begin{quote}
\textbf{Evaluation Question 4 (EQ4)}: \textit{How do the variants of the proposed approach compare to each other w.r.t.\ their computational efficiency?}
\end{quote}

Regarding this evaluation question, we expect to observe a difference in the computational cost of the proposals that use the TPE optimizer, as they rely on an iterative process that simulates and evaluates the BPS model in each iteration.
We also expect a small difference in the computational cost of the Eclipse-aware techniques when compared to the Naive approach, as the former one requires computing extra information.
Finally, regarding the ex-post and ex-ante versions of each proposal, we do not expect to observe significant differences in their computational cost, as the difference between them is only present when grouping the already discovered extraneous delays.

Henceforth, for simplicity, we will use $\mathit{Naive}$, $\mathit{EclAw}$, and $\mathit{EclAw+}$ to refer, respectively, to the Naive, Eclipse-aware, and Extrapolated eclipse-aware versions of our proposal; and $\mathit{Naive}^{TPE}$, $\mathit{EclAw}^{TPE}$, and $\mathit{EclAw+}^{TPE}$ to refer to the same versions with the TPE optimization stage.

For reproducibility purposes, all the variants of the proposed approach are implemented as a Python package publicly available (along with the code to reproduce the evaluation experiments) in GitHub.\footnote{\url{https://github.com/AutomatedProcessImprovement/extraneous-activity-delays/tree/inf-sys-3}}
Furthermore, all the employed datasets, as well as the results of the evaluation, are publicly available in Zenodo.\footnote{\url{https://doi.org/10.5281/zenodo.10435849}}

\subsection{EQ1 - Accuracy of the extraneous delay discovery\label{subsec:rq1}}

This section describes the evaluation performed to compare the proposed techniques' re-discovery accuracy (EQ1).

\vspace{10pt}
\noindent\textbf{Datasets.}
We selected a set of four BPS models (with no timer events) extracted from the examples of the book Fundamentals of Business Process Management~\cite{DBLP:books/sp/DumasRMR18}: a pharmacy prescription fulfillment process (CVS), a procure to pay process (P2P), an insurance process (INS), and a loan application process (LAP).
\tablename~\ref{tab:synthetic-log-characteristics} depicts a set of general characteristics of these processes.
We considered the original BPS models with no extraneous delays, as well as three modified versions with one, three, and five added timer events previous
to some activities (in sequential, parallel, and loop locations) to mimic the existence of extraneous activity delays.
Thus, the re-discovery evaluation in EQ1 relies on sixteen BPS models, four per process.
For each BPS model, we simulated an event log of 1,000 traces to be used as the log recording the behavior of the real process (i.e., the training log), and used the BPS models with no timers as the models to enhance.


\vspace{10pt}
\noindent\textbf{Experimental Setup.}
To evaluate the accuracy of the extraneous delay discovery, we split this evaluation question into two parts.
For the first part (EQ1.1), we preprocessed the simulated logs and computed the injected extraneous delay of each pair of causally consecutive activity instances (i.e., the exact delay that was generated from the duration distribution of the injected timer in each particular case).
Then, we run the three variants of our proposal (without the TPE optimization phase) to discover the extraneous delays associated with each pair of activity instances.
Finally, we compared each discovered extraneous delay with the injected one.
We report on the Symmetric Mean Absolute Percentage Error (SMAPE), comparing the discovered delays (forecasted values) with the injected delays (actual values).
In this computation, we only consider the transitions with injected and/or discovered extraneous delays.
The symmetric nature of the SMAPE metric gives a more robust value by fixing the asymmetry of boundlessness, while giving the percentage error allows us to perform inter-dataset comparisons.

For the second part (EQ1.2), we focus on the complete (timer event) discovery by running the three techniques with and without the TPE optimization phase, and analyzing the resulting timers w.r.t.\ the injected ones.
As measures of goodness, we use precision and recall.
In this context, a true positive stands for a discovered timer event that is present in the BPS model from which the simulated log was created, a false positive stands for a discovered timer event that is not present in the BPS model, and a false negative stands for a timer event present in the BPS model that was not discovered.
We also report on the SMAPE, comparing the mean of the discovered delays (forecasted value) with the mean of the injected extraneous delays (actual value).
We use zero as the actual value in the case of false positives, and as the forecasted value in the case of false negatives.

As configuration parameters of the approach, we used an outlier threshold of $\delta = 0.05$, only injecting timer events to those activities for which the amount of positive extraneous delays discovered was more than the 5\% of observations; a minimum short-break duration of $\lambda = 1 sec$, considering that the resources do not take breaks in between work items, to match the simulation reasoning that resources will start processing an activity as soon as they are available, and the activity enabled; a maximum scale factor of $\gamma = 10$; and set 100 iterations for the TPE hyperparameter optimization.

\begin{table}[t]
    \centering
    \footnotesize

    \caption{Characteristics of the synthetic processes used in the EQ1 and EQ2 evaluation.}
    \label{tab:synthetic-log-characteristics}

    \begin{tabular}{l r r r r}
        \toprule
        & & \multicolumn{1}{c}{Concurrent} & & \multicolumn{1}{c}{Resource} \\
        \multicolumn{1}{c}{\multirow{-2}{*}{Process}} & \multicolumn{1}{c}{\multirow{-2}{*}{Activities}} & \multicolumn{1}{c}{Activities} & \multicolumn{1}{c}{\multirow{-2}{*}{Resources}} & \multicolumn{1}{c}{Contention} \\ \toprule \toprule

        CVS   & 12  & 0  & 10  & Low      \\
        P2P   & 24  & 0  & 21  & High     \\
        INS   & 7   & 2  & 130 & Low      \\
        LAP   & 17  & 7  & 47  & High     \\ \bottomrule
    \end{tabular}
\end{table}

\vspace{10pt}
\noindent\textbf{Results.}
Regarding EQ1.1, \tablename~\ref{tab:synthetic-evaluation-individual} shows the SMAPE for the individual extraneous delay estimation.
In the datasets with no injected extraneous delays, the three techniques obtain a SMAPE of 0, meaning that they did not discover non-existent extraneous delays in any activity instance (i.e., false positives).
In the twelve datasets with injected timer events, $\mathit{EclAw+}$ presents the lower error in all the datasets, tied with $\mathit{EclAw}$ in five of them.

Regarding EQ1.2, \tablename~\ref{tab:synthetic-evaluation-complete-prec-recall} shows the precision and recall values for the complete re-discovery of the injected timer events.
In this case, the comparison is made between the three variants of our approach, with and without the TPE optimization phase.
All the techniques discovered all the injected timers (recall of 1.0), and no timers that were not injected (precision of 1.0).
Regarding the accuracy in the duration of the discovered timer delays for EQ1.2, \tablename~\ref{tab:synthetic-evaluation-complete-smape} depicts the SMAPE of the discovered timers w.r.t.\ the real (injected) ones.
All the proposals present a SMAPE of 0 in the four datasets with no injected extraneous delays, as no timers were discovered.
Regarding the other twelve datasets, $\mathit{EclAw+}$ presents the lower SMAPE errors in nine, $\mathit{EclAw+^{TPE}}$ in one, and $Naive^{TPE}$ in two.
Regarding the TPE-optimized version of each proposal, they only improved the results of their corresponding non-TPE versions in few datasets.

\begin{table}[t]
    \centering
    \footnotesize

    \caption{SMAPE of the extraneous delay estimation for each activity instance performed by $\mathit{Naive}$, $\mathit{EclAw}$, and $\mathit{EclAw+}$, in the datasets corresponding to the processes of \tablename~\ref{tab:synthetic-log-characteristics} with zero, one, three, and five injected timers (EQ1.1).
    The shadowed cells denote, for each dataset, the best result ($\pm0.01$).}
    \label{tab:synthetic-evaluation-individual}

    \setlength\tabcolsep{8pt}
    \begin{tabular}{l r r r r r r}
        \toprule
        \multicolumn{1}{c}{Dataset}  && \multicolumn{1}{c}{$\mathit{Naive}$}  && \multicolumn{1}{c}{$\mathit{EclAw}$}  && \multicolumn{1}{c}{$\mathit{EclAw+}$} \\ \toprule \toprule

        CVS\_0        && \colorcell  0.00     && \colorcell  0.00      && \colorcell 0.00      \\
        P2P\_0        && \colorcell  0.00     && \colorcell  0.00      && \colorcell 0.00      \\
        INS\_0        && \colorcell  0.00     && \colorcell  0.00      && \colorcell 0.00      \\
        LAP\_0        && \colorcell  0.00     && \colorcell  0.00      && \colorcell 0.00      \\ \midrule
        CVS\_1        &&  1.43                && 0.38                  && \colorcell 0.31      \\
        P2P\_1        &&  1.59                && \colorcell  1.08      && \colorcell 1.08      \\
        INS\_1        &&  1.05                && \colorcell  0.29      && \colorcell 0.29      \\
        LAP\_1        &&  1.70                && 1.28                  && \colorcell 1.21      \\ \midrule
        CVS\_3        &&  1.21                && 0.56                  && \colorcell 0.49      \\
        P2P\_3        &&  0.92                && \colorcell  0.68      && \colorcell 0.69      \\
        INS\_3        &&  0.81                && \colorcell  0.51      && \colorcell 0.51      \\
        LAP\_3        &&  1.41                && 1.02                  && \colorcell 0.95      \\ \midrule
        CVS\_5        &&  1.22                && 0.33                  && \colorcell 0.29      \\
        P2P\_5        &&  1.33                && 0.86                  && \colorcell 0.83      \\
        INS\_5        &&  0.74                && \colorcell  0.44      && \colorcell 0.44      \\
        LAP\_5        &&  1.45                && 1.10                  && \colorcell 1.05      \\ \bottomrule

   \end{tabular}
\end{table}

\begin{table*}[t]
    \centering
    \footnotesize

    \caption{Precision and recall of the timer event re-discovery performed by $\mathit{Naive}$, $\mathit{EclAw}$, $\mathit{EclAw+}$, $\mathit{Naive}^{TPE}$, $\mathit{EclAw}^{TPE}$, and $\mathit{EclAw+}^{TPE}$ in the datasets corresponding to the processes of \tablename~\ref{tab:synthetic-log-characteristics} with zero, one, three, and five injected timers (EQ1.2).
    The shadowed cells denote, for each dataset, the best result.}
    \label{tab:synthetic-evaluation-complete-prec-recall}

    \begin{tabular}{l r r r r r r r r r r r r}
        \toprule
        \multicolumn{1}{c}{\multirow{2}{*}{Dataset}} & \multicolumn{2}{c}{$\mathit{Naive}$}  & \multicolumn{2}{c}{$\mathit{Naive}^{TPE}$}  & \multicolumn{2}{c}{$\mathit{EclAw}$}  & \multicolumn{2}{c}{$\mathit{EclAw}^{TPE}$}  & \multicolumn{2}{c}{$\mathit{EclAw+}$}  & \multicolumn{2}{c}{$\mathit{EclAw+}^{TPE}$}  \\ \cmidrule{2-13}
          & \multicolumn{1}{c}{Prec.} & \multicolumn{1}{c}{Recall} & \multicolumn{1}{c}{Prec.} & \multicolumn{1}{c}{Recall} & \multicolumn{1}{c}{Prec.} & \multicolumn{1}{c}{Recall} & \multicolumn{1}{c}{Prec.} & \multicolumn{1}{c}{Recall} & \multicolumn{1}{c}{Prec.} & \multicolumn{1}{c}{Recall} & \multicolumn{1}{c}{Prec.} & \multicolumn{1}{c}{Recall} \\ \toprule \toprule

        CVS\_0   & \colorcell 1.00  & \colorcell 1.00  & \colorcell 1.00  & \colorcell 1.00  & \colorcell 1.00  & \colorcell 1.00  & \colorcell 1.00  & \colorcell 1.00  & \colorcell 1.00  & \colorcell 1.00  & \colorcell 1.00  & \colorcell 1.00  \\
        P2P\_0   & \colorcell 1.00  & \colorcell 1.00  & \colorcell 1.00  & \colorcell 1.00  & \colorcell 1.00  & \colorcell 1.00  & \colorcell 1.00  & \colorcell 1.00  & \colorcell 1.00  & \colorcell 1.00  & \colorcell 1.00  & \colorcell 1.00  \\
        INS\_0   & \colorcell 1.00  & \colorcell 1.00  & \colorcell 1.00  & \colorcell 1.00  & \colorcell 1.00  & \colorcell 1.00  & \colorcell 1.00  & \colorcell 1.00  & \colorcell 1.00  & \colorcell 1.00  & \colorcell 1.00  & \colorcell 1.00  \\
        LAP\_0   & \colorcell 1.00  & \colorcell 1.00  & \colorcell 1.00  & \colorcell 1.00  & \colorcell 1.00  & \colorcell 1.00  & \colorcell 1.00  & \colorcell 1.00  & \colorcell 1.00  & \colorcell 1.00  & \colorcell 1.00  & \colorcell 1.00  \\ \midrule
        CVS\_1   & \colorcell 1.00  & \colorcell 1.00  & \colorcell 1.00  & \colorcell 1.00  & \colorcell 1.00  & \colorcell 1.00  & \colorcell 1.00  & \colorcell 1.00  & \colorcell 1.00  & \colorcell 1.00  & \colorcell 1.00  & \colorcell 1.00  \\
        P2P\_1   & \colorcell 1.00  & \colorcell 1.00  & \colorcell 1.00  & \colorcell 1.00  & \colorcell 1.00  & \colorcell 1.00  & \colorcell 1.00  & \colorcell 1.00  & \colorcell 1.00  & \colorcell 1.00  & \colorcell 1.00  & \colorcell 1.00  \\
        INS\_1   & \colorcell 1.00  & \colorcell 1.00  & \colorcell 1.00  & \colorcell 1.00  & \colorcell 1.00  & \colorcell 1.00  & \colorcell 1.00  & \colorcell 1.00  & \colorcell 1.00  & \colorcell 1.00  & \colorcell 1.00  & \colorcell 1.00  \\
        LAP\_1   & \colorcell 1.00  & \colorcell 1.00  & \colorcell 1.00  & \colorcell 1.00  & \colorcell 1.00  & \colorcell 1.00  & \colorcell 1.00  & \colorcell 1.00  & \colorcell 1.00  & \colorcell 1.00  & \colorcell 1.00  & \colorcell 1.00  \\ \midrule
        CVS\_3   & \colorcell 1.00  & \colorcell 1.00  & \colorcell 1.00  & \colorcell 1.00  & \colorcell 1.00  & \colorcell 1.00  & \colorcell 1.00  & \colorcell 1.00  & \colorcell 1.00  & \colorcell 1.00  & \colorcell 1.00  & \colorcell 1.00  \\
        P2P\_3   & \colorcell 1.00  & \colorcell 1.00  & \colorcell 1.00  & \colorcell 1.00  & \colorcell 1.00  & \colorcell 1.00  & \colorcell 1.00  & \colorcell 1.00  & \colorcell 1.00  & \colorcell 1.00  & \colorcell 1.00  & \colorcell 1.00  \\
        INS\_3   & \colorcell 1.00  & \colorcell 1.00  & \colorcell 1.00  & \colorcell 1.00  & \colorcell 1.00  & \colorcell 1.00  & \colorcell 1.00  & \colorcell 1.00  & \colorcell 1.00  & \colorcell 1.00  & \colorcell 1.00  & \colorcell 1.00  \\
        LAP\_3   & \colorcell 1.00  & \colorcell 1.00  & \colorcell 1.00  & \colorcell 1.00  & \colorcell 1.00  & \colorcell 1.00  & \colorcell 1.00  & \colorcell 1.00  & \colorcell 1.00  & \colorcell 1.00  & \colorcell 1.00  & \colorcell 1.00  \\ \midrule
        CVS\_5   & \colorcell 1.00  & \colorcell 1.00  & \colorcell 1.00  & \colorcell 1.00  & \colorcell 1.00  & \colorcell 1.00  & \colorcell 1.00  & \colorcell 1.00  & \colorcell 1.00  & \colorcell 1.00  & \colorcell 1.00  & \colorcell 1.00  \\
        P2P\_5   & \colorcell 1.00  & \colorcell 1.00  & \colorcell 1.00  & \colorcell 1.00  & \colorcell 1.00  & \colorcell 1.00  & \colorcell 1.00  & \colorcell 1.00  & \colorcell 1.00  & \colorcell 1.00  & \colorcell 1.00  & \colorcell 1.00  \\
        INS\_5   & \colorcell 1.00  & \colorcell 1.00  & \colorcell 1.00  & \colorcell 1.00  & \colorcell 1.00  & \colorcell 1.00  & \colorcell 1.00  & \colorcell 1.00  & \colorcell 1.00  & \colorcell 1.00  & \colorcell 1.00  & \colorcell 1.00  \\
        LAP\_5   & \colorcell 1.00  & \colorcell 1.00  & \colorcell 1.00  & \colorcell 1.00  & \colorcell 1.00  & \colorcell 1.00  & \colorcell 1.00  & \colorcell 1.00  & \colorcell 1.00  & \colorcell 1.00  & \colorcell 1.00  & \colorcell 1.00  \\ \bottomrule
    \end{tabular}

\end{table*}

\begin{table*}[t]
    \centering
    \footnotesize

    \caption{SMAPE of the timer event re-discovery performed by $\mathit{Naive}$, $\mathit{EclAw}$, $\mathit{EclAw+}$, $\mathit{Naive}^{TPE}$, $\mathit{EclAw}^{TPE}$, and $\mathit{EclAw+}^{TPE}$ in the datasets corresponding to the processes of \tablename~\ref{tab:synthetic-log-characteristics} with zero, one, three, and five injected timers (EQ1.2).
    The shadowed cells denote, for each dataset, the best result ($\pm0.01$).}
    \label{tab:synthetic-evaluation-complete-smape}
    \begin{tabular}{l r r r r r r r r r r r r}
        \toprule
        \multicolumn{1}{c}{Dataset} && \multicolumn{1}{c}{$\mathit{Naive}$} && \multicolumn{1}{c}{$\mathit{Naive}^{TPE}$} && \multicolumn{1}{c}{$\mathit{EclAw}$} && \multicolumn{1}{c}{$\mathit{EclAw}^{TPE}$} && \multicolumn{1}{c}{$\mathit{EclAw+}$} && \multicolumn{1}{c}{$\mathit{EclAw+}^{TPE}$} \\ \toprule \toprule

        CVS\_0   && \colorcell 0.00  &&  \colorcell 0.00  &&  \colorcell 0.00  &&  \colorcell 0.00  &&  \colorcell 0.00  &&  \colorcell 0.00  \\
        P2P\_0   && \colorcell 0.00  &&  \colorcell 0.00  &&  \colorcell 0.00  &&  \colorcell 0.00  &&  \colorcell 0.00  &&  \colorcell 0.00  \\
        INS\_0   && \colorcell 0.00  &&  \colorcell 0.00  &&  \colorcell 0.00  &&  \colorcell 0.00  &&  \colorcell 0.00  &&  \colorcell 0.00  \\
        LAP\_0   && \colorcell 0.00  &&  \colorcell 0.00  &&  \colorcell 0.00  &&  \colorcell 0.00  &&  \colorcell 0.00  &&  \colorcell 0.00  \\ \midrule
        CVS\_1   && 1.36             &&             0.84  &&             0.30  &&             0.51  &&  \colorcell 0.02  &&             0.70  \\
        P2P\_1   && 1.51             &&  \colorcell 0.34  &&             1.02  &&             0.69  &&             0.37  &&             1.37  \\
        INS\_1   && 1.01             &&  \colorcell 0.08  &&             0.26  &&             0.28  &&             0.24  &&             0.29  \\
        LAP\_1   && 1.65             &&             0.34  &&             1.16  &&             0.92  &&  \colorcell 0.11  &&             1.55  \\ \midrule
        CVS\_3   && 1.12             &&             0.48  &&             0.44  &&             0.68  &&  \colorcell 0.12  &&             0.51  \\
        P2P\_3   && 0.73             &&             1.07  &&             0.54  &&             1.04  &&  \colorcell 0.37  &&             1.39  \\
        INS\_3   && 0.74             &&             0.54  &&  \colorcell 0.52  &&             0.61  &&  \colorcell 0.51  &&             0.94  \\
        LAP\_3   && 1.16             &&             1.10  &&             0.84  &&             1.07  &&  \colorcell 0.12  &&             1.57  \\ \midrule
        CVS\_5   && 1.17             &&             0.48  &&  \colorcell 0.26  &&             0.74  &&             0.28  &&  \colorcell 0.25  \\
        P2P\_5   && 1.17             &&             0.91  &&             0.65  &&             0.40  &&  \colorcell 0.28  &&             1.48  \\
        INS\_5   && 0.69             &&             0.53  &&  \colorcell 0.42  &&             0.85  &&  \colorcell 0.43  &&             0.85  \\
        LAP\_5   && 1.30             &&             1.18  &&             0.92  &&             1.44  &&  \colorcell 0.37  &&             1.24  \\ \bottomrule
    \end{tabular}

\end{table*}


\vspace{10pt}
\noindent\textbf{Discussion.}
Regarding the accuracy of the individual extraneous delay estimation (EQ1.1), the results are as expected.
In the scenarios with no injected delays, all the waiting time is explained by resource contention and resource unavailability, and none of the techniques detected non-existent extraneous delays.
In the scenarios with injected timers, $\mathit{EclAw+}$ presents the lower SMAPE values in all datasets, closely followed by $\mathit{EclAw}$ with an average difference in the SMAPE of $0.03$, while $\mathit{Naive}$ presents the worst results.
$\mathit{Naive}$ follows the most conservative approach, only considering as extraneous delay the latest interval of waiting time when the activity could have been executed (see \definitionname~\ref{def:naive-extraneous-delay}).
As anticipated, the presence of eclipses (see \sectionname~\ref{subsubsec:eclipse-aware}) results in a high error in the duration of the extraneous delay.
On the other hand, the heuristic followed by $\mathit{EclAw}$ drastically reduces the impact of the eclipses.
Finally, $\mathit{EclAw+}$ goes one step forward and extrapolates the extraneous delay to reduce the error in average (see \definitionname~\ref{def:complex-extrapolated-extraneous-delay}).
This adjustment improves the results in most of the datasets.

Interestingly, the impact of the resource contention is also visible in the results of $\mathit{EclAw}$ and $\mathit{EclAw+}$.
The error is higher in the datasets of the processes with more resource contention (P2P and LAP).
A higher resource contention increases the probability of other activity instances to eclipse the extraneous delays, increasing the error in the estimation.

Regarding the accuracy in the complete discovery of timers (EQ1.2), the results also follow an expected structure given the analysis of EQ1.1.
All the proposals successfully discovered the presence of all the injected timer events. None of the approaches discovered non-existent timer events. In other words, all proposals achieved a precision and recall of 1.0.
With respect to the duration of the discovered timer events, $\mathit{EclAw+}$ shows the best results in most of the datasets, due to the reasons commented above.
It is worth mentioning that, although $\mathit{Naive}$ presents the worst results in \textit{P2P\_1} and \textit{INS\_1}, its TPE-optimized version achieved the lower SMAPE among all techniques.
In some situations, a worse approximation may lead the TPE optimization through a better path, ending in better results.
However, as seen in the results, this is not common.

Surprisingly, the TPE-optimized version of each proposal only improved the results (w.r.t. the non optimized version) in a few of the datasets.
This is explained by the nature of this experiment.
As the real extraneous delays are unknown to the approaches, the TPE stage is designed to optimize a distance measure between the distribution of events produced by the discovered simulation model and the original event log.
In some situations, the path taken by the optimizer leads to a better solution w.r.t.\ this metric, but not w.r.t.\ the real duration distribution of each timer.
Furthermore, to avoid overfitting, the TPE optimization process discovers the delays with a sample of the training log (50\% of all the events), and uses the rest to evaluate (validation set) the solution in each iteration.
This partitioning process can negatively impact the extraneous delay estimation and result in a slightly higher error in the initial estimation.

\subsection{EQ2 - ex-post vs ex-ante discovery}

This section describes the evaluation performed to analyze the differences in the discovery between the ex-post and ex-ante options of the proposed approaches (EQ2).

\vspace{10pt}
\noindent\textbf{Datasets.}
We used the same set of four BPS models employed in \sectionname~\ref{subsec:rq1}.
Then, for each of these processes, we added five timer events to the BPS model.
With the purpose of studying the approach behavior under different scenarios, we placed the timers in distinct locations (between two sequential activities, after one activity and before a split gateway, in each of the branches of a parallel structure, as part of a loop, etc.).
Thus, EQ2 relies on four BPS models (one per process), each one with five timer events to simulate extraneous delays.
For each BPS model, we simulated an event log of 1,000 traces to be used as the log recording the behavior of the real process (i.e., the training log), and used the BPS models with no timers as the models to enhance.

\vspace{10pt}
\noindent\textbf{Experimental Setup.}
We are interested in qualitatively analyzing the differences between the ex-post and the ex-ante variants of our proposal.
For this reason, we run both $\mathit{Naive}$ and $\mathit{EclAw+}$ (ex-post and ex-ante configurations) to discover the injected timer delays with each event log.
We qualitatively analyzed if there is any difference between the two proposals, and, more importantly, how the ex-post and ex-ante configurations affect the timers' discovery.
We decided not to evaluate $\mathit{EclAw}$ in EQ2 because this approach consists of the same heuristics as $\mathit{EclAw+}$ -- both approaches would discover the same timers, but with different durations --, and we are only interested in analyzing how the discovery of timer events changes depending on the timer-placement configuration.
For the same reason, we excluded the TPE-optimized versions of the proposals, as the TPE phase is only designed to tune the duration of the timers.
As configuration parameters of the approach, we used the same values as for EQ1.

\begin{figure*}[t]
    \begin{minipage}{0.13\textwidth}
    	\centering
	    \subfloat[]{
            \includegraphics[width=0.99\textwidth]{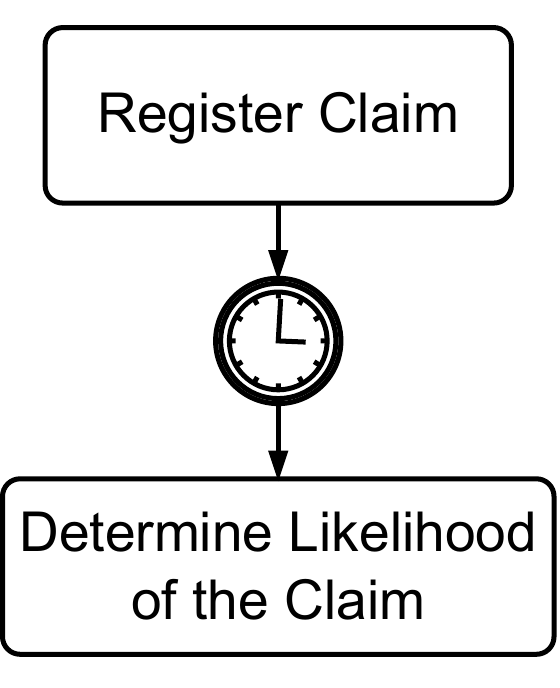}
	        \label{subfig:befaft-sequence}
	    }
    \end{minipage}
    \hfill
    \begin{minipage}{0.23\textwidth}
    	\centering
	    \subfloat[]{
            \includegraphics[width=0.99\textwidth]{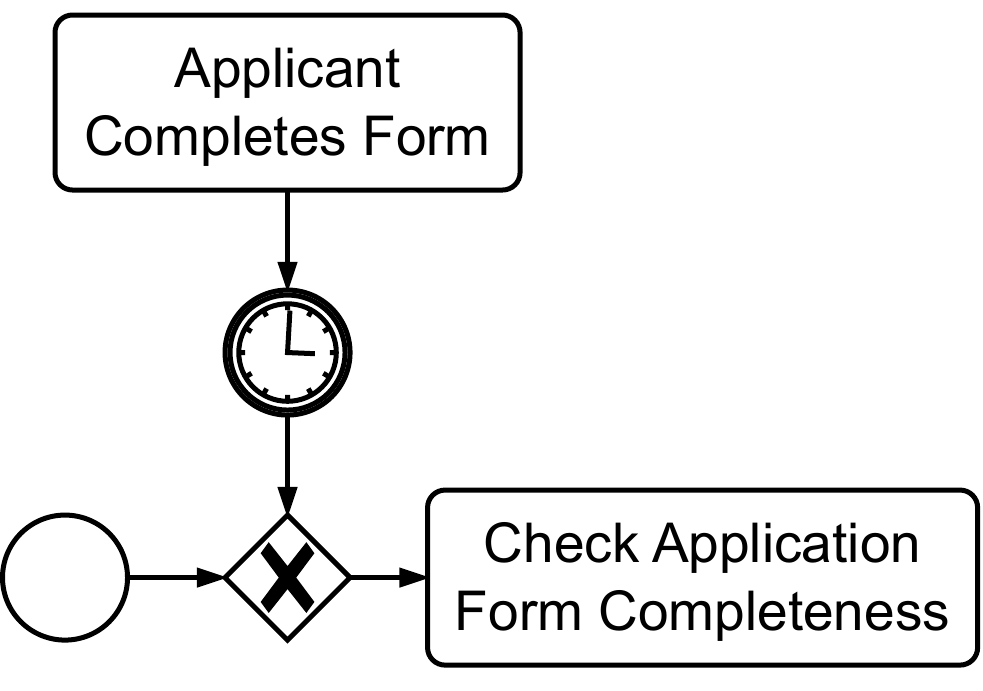}
	        \label{subfig:befaft-xor-join}
	    }
    \end{minipage}
    \hfill
    \begin{minipage}{0.27\textwidth}
    	\centering
	    \subfloat[]{
            \includegraphics[width=0.99\textwidth]{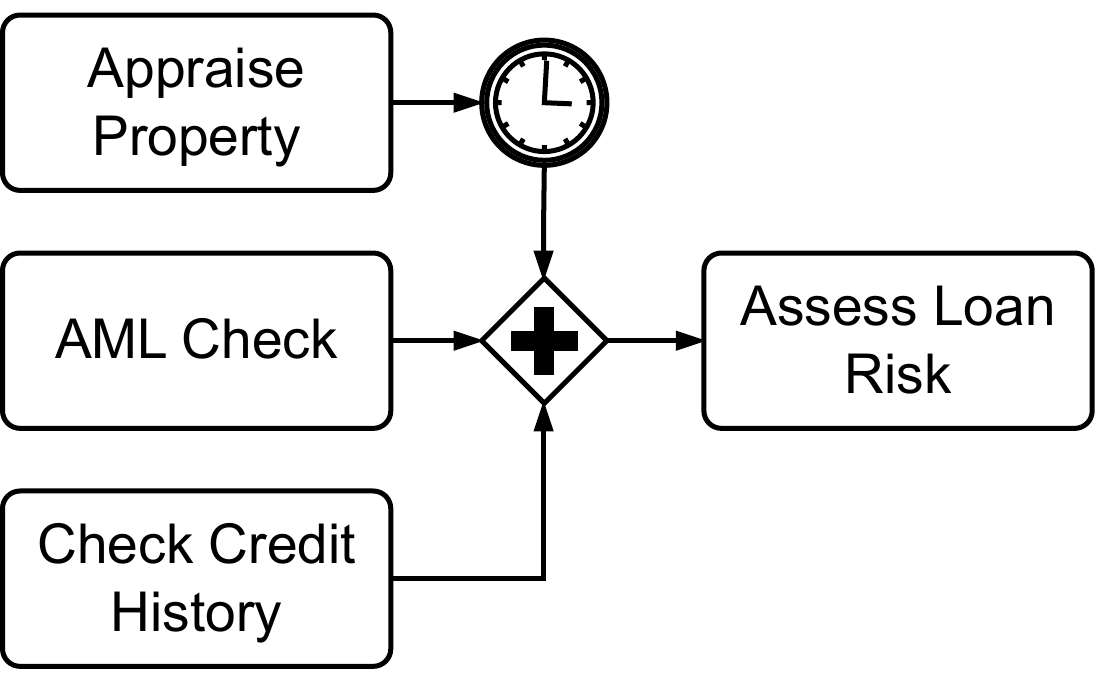}
	        \label{subfig:befaft-and-join-single}
	    }
    \end{minipage}
    \hfill
    \begin{minipage}{0.22\textwidth}
    	\centering
	    \subfloat[]{
            \includegraphics[width=0.99\textwidth]{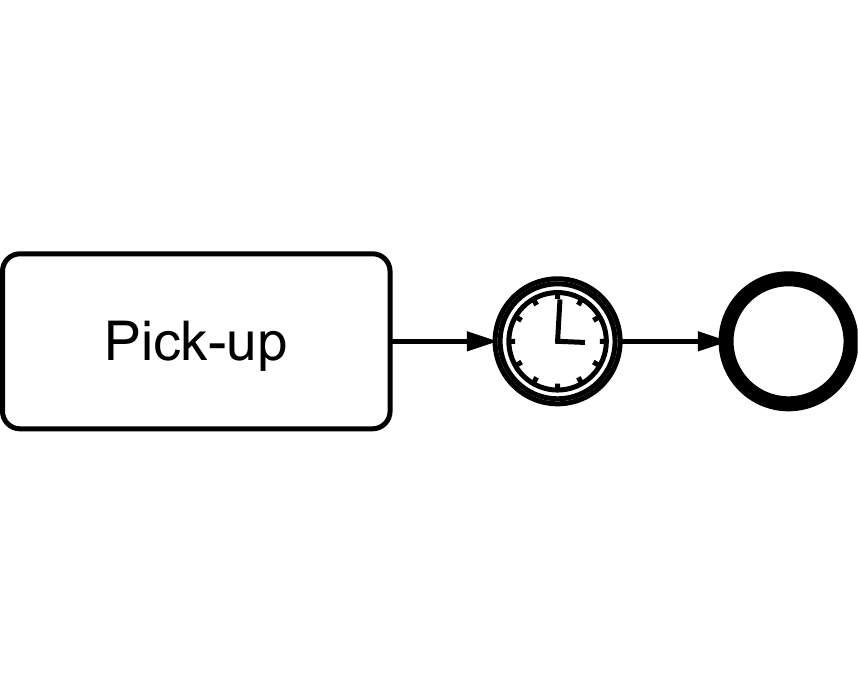}
	        \label{subfig:befaft-end-event}
	    }
    \end{minipage}
    \hfill

    \caption{Four situations from the datasets used in EQ2 to exemplify the impact of considering the delay as an ex-post cause of the enabling activity, or ex-ante cause of the executed activity.}
    \label{fig:qualitative-analysis-timer-placement}
\end{figure*}

\vspace{10pt}
\noindent\textbf{Results.}
Regarding the differences in how the extraneous delays are discovered depending on their location and of the ex-post or ex-ante configuration, \figurename~\ref{fig:qualitative-analysis-timer-placement} depicts four different situations observed in this evaluation.
In all the cases, both techniques ($\mathit{Naive}$ and $\mathit{EclAw+}$) discovered the same timer events.
However, the ex-post and ex-ante configurations differed depending on the timer location.
The simplest scenario is shown in \figurename~\ref{subfig:befaft-sequence}, where the timer is placed in between a sequence of two activities.
In this case, the ex-post configuration associates the timer to the source activity (''Register Claim``), while the ex-ante configuration associates it to the target activity (''Determine Likelihood of the Claim``).
Nevertheless, the duration of the discovered extraneous delays is similar in both configurations.

\figurename~\ref{subfig:befaft-xor-join} shows another example where the timer event is placed before an XOR gateway.
In these situations, when the configuration of the approach matches the placement of the timer (ex-post in the example of the figure), the timer is discovered without alterations.
However, in the case of the opposite configuration (ex-ante in the example of the figure), the timer gets scattered among the gateway branches.
In the example of the figure, the discovery of the timer in this latter configuration will place it previous to ''Check Application Form Completeness`` and, thus, simulate extraneous delays when the activity is executed after ''Applicant Completes Form``, but also after the start of the trace.
In the same way, if the XOR gateway has multiple outgoing edges, the ex-ante configuration would scatter the delays through each of the branches, placing a delay in each of them.
The same behavior, but benefiting the ex-ante configuration, occurs if the timer is after the XOR gateway.

A similar situation is shown in \figurename~\ref{subfig:befaft-and-join-single}, where the gateway connected to the timer is an AND gateway instead of an XOR gateway.
The results in this scenario are the same as the previous one when the opposite timer-placement configuration is used in this scenario.
However, AND gateways present an additional case.
In an ex-post configuration, the extraneous delay is attributed to the source activity instance of each pair.
Thus, if the other parallel activities finish their execution before the extraneous delay finishes, the delay will be imputed to the last ending activity instance (i.e., the one that will be present in the pair of causally consecutive activity instances).
However, if the combination of ''Appraise Property`` and the extraneous delay is always shorter than one of the other parallel activities, no extraneous delay will be registered, as ''Assess Loan Risk`` will always start after the longer parallel activity ends.
Thus, for an ex-post configuration, this situation can lead to either no discovered delays, or any of the activities in the parallel structure being potentially imputed with extraneous delay.

Finally, \figurename~\ref{subfig:befaft-end-event} shows a situation where no delay is discovered.
Typically, event logs do not contain information related to the start or end of each case (we consider, respectively, the first start time and last end times).
For this reason, delays preceding the end event, or succeeding the start event, are not discovered (given the information available in the event log, the trace finished when ''Pick-up`` ended its execution).

\vspace{10pt}
\noindent\textbf{Discussion.}
This evaluation reported many differences in the discovery of the injected timers depending on the timer-placement configuration.
However, these differences lie in if the configuration is aligned or not with the nature of the timer (e.g., if the extraneous delay occurs after an activity, the ex-post configuration will generally perform better).
For this reason, we conclude that the differences between the timer-placement configuration are conceptual, depending on the domain and individual aspects of the process, rather than performance-related.

\subsection{EQ3 \& EQ4 - Accuracy of the enhanced simulation and computational efficiency\label{subsec:rq3}}

This section describes the evaluation performed to analyze the improvement of our proposal in the temporal accuracy of a BPS model in real-life scenarios (EQ3), as well as the analysis of the computational efficiency of each of the proposed techniques (EQ4).

\vspace{10pt}
\noindent\textbf{Datasets.}
We selected three real-life event logs with both start and end timestamps, resource information, and low multitasking.\footnote{We selected these three real-life datasets because they meet the necessary characteristics for BPS: activity instances with known duration (i.e., start and end timestamps) and resource information.
In addition, these logs have a low level of multitasking which, as described in \sectionname~\ref{sec:approach}, constitutes the best scenario for testing our proposal.}
One of them is a log of an academic credentials' management process (AC\_CRE), containing a high number of resources exhibiting low participation in the process.
The other two logs come from the Business Process Intelligence Challenges (BPIC) of 2012~\cite{vanDongen_BPIC12} and 2017~\cite{vanDongen_BPIC17}.
We preprocessed the event log from BPIC 2012, retaining only the events corresponding to activities performed by human resources (i.e., only activity instances that have a duration), and the event log from BPIC 2017 by following the recommendations reported by the winning teams participating in the competition (\url{https://www.win.tue.nl/bpi/doku.php?id=2017:challenge}).
Then, in order to avoid data leakage in the evaluation of our approach, we split each event log into two sets (the training and the testing set).
For each dataset, we selected two temporal disjoint intervals (i.e., non-overlapping) with similar trace and event intensity (avoiding the concept change in the activity instances intensity in the BPIC 2017), and retained only the traces that were fully contained in them.
In this way, no events in the training set overlap with events in the test set.
\tablename~\ref{tab:real-log-characteristics} shows the characteristics of the two training (BPIC12TR and BPIC17TR) and two test (BPIC12TE and BPIC17TE) event logs.
To obtain the BPS model to enhance, we used SIMOD~\cite{DBLP:journals/dss/CamargoDG20} to automatically discover a BPS model from the training partitions.

\begin{table*}[t]
    \centering
    \footnotesize

    \caption{Characteristics of the real-life event logs used in the evaluation of EQ3 and EQ4.}
    \label{tab:real-log-characteristics}

    \begin{tabular}{l r r r r r}
        \toprule
                     \multicolumn{1}{c}{Event log} & \multicolumn{1}{c}{Traces} & \multicolumn{1}{c}{Activity instances} & \multicolumn{1}{c}{Variants} & \multicolumn{1}{c}{Activities} & \multicolumn{1}{c}{Resources} \\ \toprule \toprule
        AC\_CRE\_TR  & 398      & 1,945     & 54      & 16  & 306     \\
        AC\_CRE\_TE  & 398      & 1,788     & 35      & 16  & 281     \\ \midrule
        BPIC12\_TR   & 3,030    & 16,338    & 735     & 6   & 47      \\
        BPIC12\_TE   & 2,976    & 18,568    & 868     & 6   & 53      \\ \midrule
        BPIC17\_TR   & 7,402    & 53,332    & 1,843   & 7   & 105     \\
        BPIC17\_TE   & 7,376    & 52,010    & 1,830   & 7   & 113     \\ \bottomrule
    \end{tabular}
\end{table*}


\vspace{10pt}
\noindent\textbf{Experimental Setup.}
To analyze the improvement of the different variants of our proposal, we compared the quality of the BPS model discovered by an automated BPS model discovery technique (SIMOD), with the quality of the BPS models enhanced by $\mathit{Naive}$, $\mathit{EclAw+}$, $\mathit{Naive^{TPE}}$, and $\mathit{EclAw+^{TPE}}$.\footnote{$\mathit{EclAw+}$ is an extension of $\mathit{EclAw}$ that improves the average accuracy in the duration of the estimated delays (see \sectionname~\ref{subsubsec:eclipse-aware-extrapolation} and \sectionname~\ref{subsec:rq1}). Furthermore, the computational complexity of both techniques is similar, as the only difference is an additional division when computing the delay duration. For these reasons, we only analyze the performance of $\mathit{EclAw+}$ in this evaluation question.}
To assess the quality of the BPS models, we followed the evaluation framework presented in~\cite{DBLP:conf/bpm/ChapelaCampaBBDKS23}.
Accordingly, for each process, we generated a collection of ten logs simulated with the BPS model under evaluation, and compared them to the testing event log (i.e., the system that the model aims to mimic).
In previous work, the distance between two event logs is made along two dimensions: the control-flow and the temporal dimensions~\cite{DBLP:journals/dss/CamargoDG20,DBLP:journals/peerj-cs/CamargoDR21,DBLP:conf/bpm/ChapelaCampaBBDKS23}.
Given that the modeling of extraneous delays does not alter the control-flow of the BPS model, we hereby focus on the temporal dimension.
Specifically, we compared each simulated log against the corresponding test log in two ways: \textit{i)} by comparing the distribution of events within each trace (i.e., their distribution w.r.t.\ the start of the trace); and \textit{ii)} by comparing the distribution of cycle times of their traces.

\begin{table*}[b]
    \centering
    \footnotesize
    \caption{Relative Event Distribution distance values (mean and confidence interval) between each real-life process and the evaluated BPS models (the one discovered by SIMOD, and ones enhanced with our proposal).
    The shadowed cells denote, for each dataset, the best results out of all the techniques.}
    \label{tab:real-life-results}

    \begin{tabular}{l r r r r r}
        \toprule
        \multicolumn{1}{c}{\multirow{2}{*}{Dataset}} & \multicolumn{1}{c}{\multirow{2}{*}{SIMOD}} & \multicolumn{2}{c}{$\mathit{Naive}$}                                 & \multicolumn{2}{c}{$\mathit{Naive^{TPE}}$} \\ \cmidrule{3-6}
        &  & \multicolumn{1}{c}{ex-post} & \multicolumn{1}{c}{ex-ante} & \multicolumn{1}{c}{ex-post} & \multicolumn{1}{c}{ex-ante} \\ \toprule \toprule

        AC\_CRE      & 77.21($\pm$0.35)  & 77.20($\pm$0.46)  & 77.28($\pm$0.40)  & 75.52($\pm$0.35)  & 75.75($\pm$0.37)  \\
        BPIC12       & 178.55($\pm$0.01) & 171.02($\pm$0.32) & 167.94($\pm$0.37) & 148.01($\pm$1.26) & 137.87($\pm$1.37) \\
        BPIC17       & 183.55($\pm$0.02) & 183.05($\pm$0.03) & 183.10($\pm$0.02) & 179.66($\pm$0.03) & 180.15($\pm$0.05) \\ \bottomrule
    \end{tabular}

    \vspace{2pt}

    \begin{tabular}{l rrrrr r r r r}
        \multicolumn{1}{c}{\multirow{2}{*}{Dataset}} &&&&&  & \multicolumn{2}{c}{$\mathit{EclAw+}$}                        & \multicolumn{2}{c}{$\mathit{EclAw+^{TPE}}$}                  \\ \cmidrule{7-10}
        & &&&&& \multicolumn{1}{c}{ex-post} & \multicolumn{1}{c}{ex-ante} & \multicolumn{1}{c}{ex-post} & \multicolumn{1}{c}{ex-ante} \\ \toprule \toprule

        AC\_CRE     & &&&& & 41.51($\pm$2.22)  & \colorcell34.15($\pm$1.53)  & 41.02($\pm$1.84)  &  \phantom{1}\colorcell35.89($\pm$0.93) \\
        BPIC12      & &&&& & 143.89($\pm$0.67) & 126.95($\pm$0.66) & 96.16($\pm$1.36) &  \phantom{1}\colorcell57.94($\pm$2.06) \\
        BPIC17      & &&&& & 119.33($\pm$0.53) & 106.86($\pm$0.54)  & 41.70($\pm$0.30)  &  \phantom{1}\colorcell26.03($\pm$0.81) \\ \bottomrule
    \end{tabular}

\end{table*}

\begin{table*}[t]
    \centering
    \footnotesize
    \caption{Runtime values (in seconds) for the different versions of our proposal.}
    \label{tab:real-life-runtimes}

    \begin{tabular}{l r r r r r r r r}
        \toprule
        \multicolumn{1}{c}{\multirow{2}{*}{Dataset}} & \multicolumn{2}{c}{$\mathit{Naive}$}                                 & \multicolumn{2}{c}{$\mathit{Naive^{TPE}}$}                           & \multicolumn{2}{c}{$\mathit{EclAw+}$}                        & \multicolumn{2}{c}{$\mathit{EclAw+^{TPE}}$}                  \\ \cmidrule{2-9}
        &  \multicolumn{1}{c}{ex-post} & \multicolumn{1}{c}{ex-ante} & \multicolumn{1}{c}{ex-post} & \multicolumn{1}{c}{ex-ante} & \multicolumn{1}{c}{ex-post} & \multicolumn{1}{c}{ex-ante} & \multicolumn{1}{c}{ex-post} & \multicolumn{1}{c}{ex-ante} \\ \toprule \toprule

        AC\_CRE      & 0.04 & 5.76  & 206.18   & 210.34   & 4.71   & 4.47   & 218.50   & 212.23   \\
        BPIC12       & 3.31 & 22.82 & 1,445.54 & 1,438.67 & 42.08  & 41.36  & 1,496.50 & 1,529.73 \\
        BPIC17       & 8.31 & 70.55 & 3,642.16 & 3,565.30 & 156.38 & 160.22 & 3,381.50 & 3,817.01 \\ \bottomrule
    \end{tabular}

\end{table*}

To compare the distribution of events within each trace, we use the Relative Event Distribution (RED) distance measure proposed in~\cite{DBLP:conf/bpm/ChapelaCampaBBDKS23}.
This measure first transforms each event (both start and end events of each activity instance) in the log into its temporal difference w.r.t.\ the start of its trace.
Then, it discretizes these values by grouping them into bins of one hour (e.g., all the events recorded between 3600 and 7199 seconds since the start of their trace go to the second bin).
The result is a discretized time series with the distribution of recorded events w.r.t.\ the start of their trace.
Finally, the RED distance corresponds to the Earth Movers' Distance between the two discretized distributions.
The Earth Movers' Distance interprets the two histograms as piles of ``dirt'', where each bin is a pile, and measures the amount of dirt that has to be moved from one histogram to obtain the other, penalizing the movements by their distance.
In this way, a higher value denotes a higher distance (i.e., lower similarity).
Importantly, the magnitude of the EMD depends on the dataset (e.g., larger datasets will naturally give rise to larger EMD values).
Accordingly, we use the EMD to make intra-dataset comparisons only (i.e., to measure the relative performance of multiple techniques within the same dataset).

To compare the distribution of cycle times, we report the minimum, first quartile, median, average, third quartile, and maximum values of the simulated logs (as the average of the ten runs), and compare them with the same values in the testing log.
Finally, to also analyze the computational efficiency of each technique, we report the runtime of each of the proposals for each dataset.
As configuration parameters of the approach, we used the same values as for EQ1.2.

\vspace{10pt}
\noindent\textbf{Results.}
Regarding the distribution of events within each process trace, \tablename~\ref{tab:real-life-results} shows the RED distance values of the BPS model discovered by SIMOD, and the BPS models enhanced with the different variants of our proposal.
All the enhanced BPS models present, in the three datasets, better results than the BPS model discovered by SIMOD.
With respect to the ex-post and ex-ante versions of each proposal, the latter one present better results in $\mathit{EclAw+}$ and $\mathit{EclAw+^{TPE}}$, and similar ones for $\mathit{Naive}$ and $\mathit{Naive^{TPE}}$.
Focusing on each technique, $\mathit{Naive}$ only presents a small improvement w.r.t.\ the BPS model discovered by SIMOD in the BPIC12 dataset.
However, when running the technique combined with the TPE optimizer ($\mathit{Naive^{TPE}}$), the improvement is visible in the three datasets.
These results are overcome by the BPS models enhanced with $\mathit{EclAw+}$.
Finally, the lowest distances are presented by $\mathit{EclAw+^{TPE}}$.

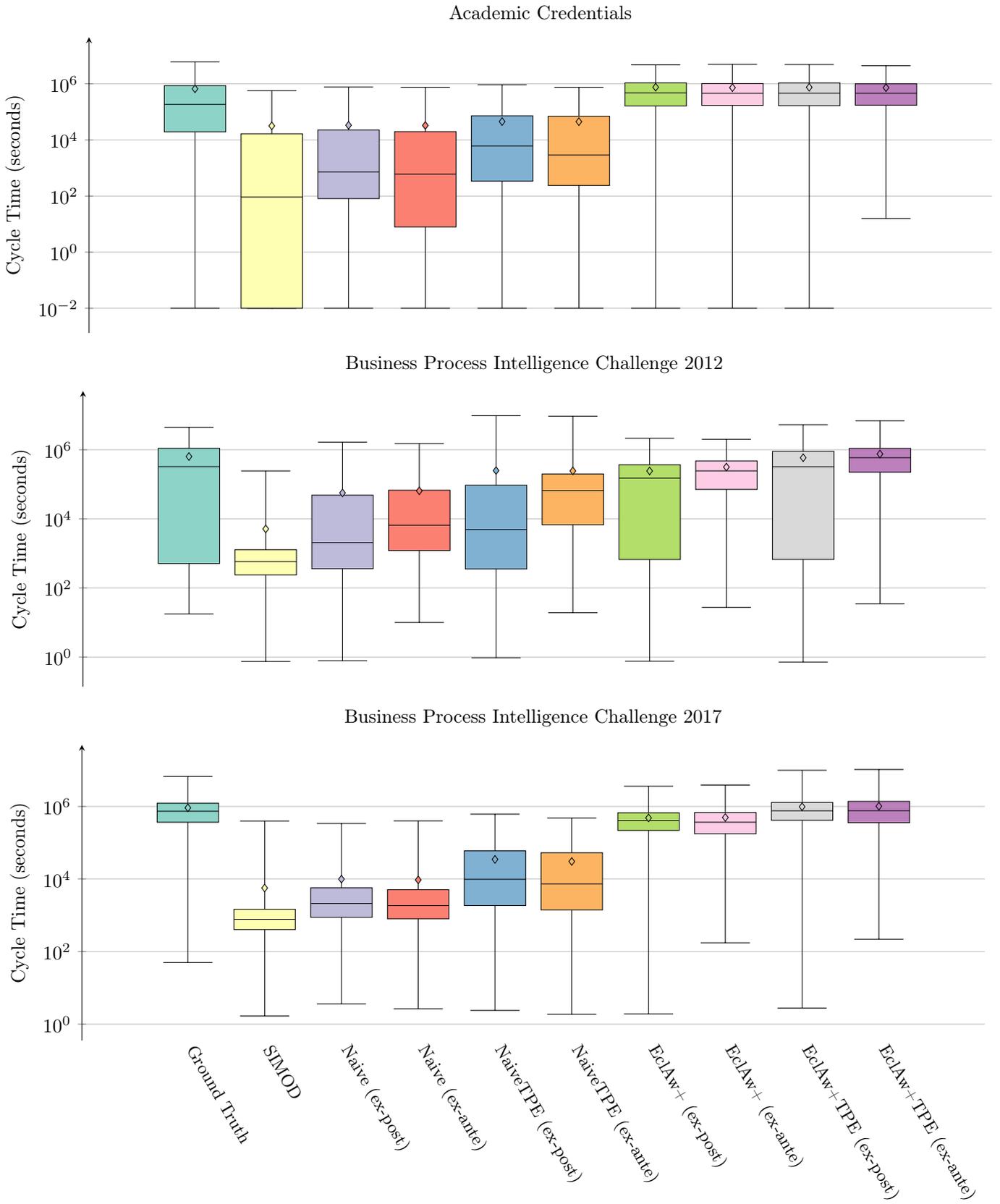
\begin{figure*}[h!]
    \centering
    \begin{tikzpicture}
    	\pgfplotstableread[col sep=comma]{data/cycle-time-academic.csv}\csvdata
        \pgfplotstablegetrowsof{\csvdata}
        \pgfmathtruncatemacro\TotalRows{\pgfplotsretval-1}
    	\begin{axis}[
    		boxplot/draw direction = y,
            title = Academic Credentials,
    		x axis line style = {opacity=0},
    		axis x line* = bottom,
    		axis y line = left,
            height = 7cm,
            width = 18cm,
    		enlarge y limits,
    		ymajorgrids,
    		xtick = {1, 2, 3, 4, 5, 6, 7, 8, 9, 10},
    		xticklabel style = {align=center, font=\small, rotate=300, anchor=west},
            xticklabels = {},
    		xtick style = {draw=none}, 
    		ylabel = {Cycle Time (seconds)},
            ymode=log
    	]
            \pgfplotsinvokeforeach{0,...,\TotalRows}
            {
              \addplot+[
                  mark = *,
                  boxplot prepared from table={
                    table=\csvdata,
                    row=#1,
                    lower whisker=min_mean,
                    lower quartile=q1_mean,
                    median=median_mean,
                    average=mean_mean,
                    upper quartile=q3_mean,
                    upper whisker=max_mean
                  },
                  boxplot prepared,
                  fill,
                  draw=black
              ]
              coordinates {};
            }
    	\end{axis}
    \end{tikzpicture}

    \begin{tikzpicture}
    	\pgfplotstableread[col sep=comma]{data/cycle-time-bpic12.csv}\csvdata
        \pgfplotstablegetrowsof{\csvdata}
        \pgfmathtruncatemacro\TotalRows{\pgfplotsretval-1}
    	\begin{axis}[
    		boxplot/draw direction = y,
            title = Business Process Intelligence Challenge 2012,
    		x axis line style = {opacity=0},
    		axis x line* = bottom,
    		axis y line = left,
            height = 7cm,
            width = 18cm,
    		enlarge y limits,
    		ymajorgrids,
    		xtick = {1, 2, 3, 4, 5, 6, 7, 8, 9, 10},
    		xticklabel style = {align=center, font=\small, rotate=300, anchor=west},
            xticklabels = {},
    		xtick style = {draw=none}, 
    		ylabel = {Cycle Time (seconds)},
            ymode=log
    	]
            \pgfplotsinvokeforeach{0,...,\TotalRows}
            {
              \addplot+[
              boxplot prepared from table={
                table=\csvdata,
                row=#1,
                lower whisker=min_mean,
                lower quartile=q1_mean,
                median=median_mean,
                average=mean_mean,
                upper quartile=q3_mean,
                upper whisker=max_mean
              },
              boxplot prepared,
              fill,
              draw=black
              ]
              coordinates {};
            }
    	\end{axis}
    \end{tikzpicture}

    \begin{tikzpicture}
    	\pgfplotstableread[col sep=comma]{data/cycle-time-bpic17.csv}\csvdata
        \pgfplotstablegetrowsof{\csvdata}
        \pgfmathtruncatemacro\TotalRows{\pgfplotsretval-1}
    	\begin{axis}[
    		boxplot/draw direction = y,
            title = Business Process Intelligence Challenge 2017,
    		x axis line style = {opacity=0},
    		axis x line* = bottom,
    		axis y line = left,
            height = 7cm,
            width = 18cm,
    		enlarge y limits,
    		ymajorgrids,
    		xtick = {1, 2, 3, 4, 5, 6, 7, 8, 9, 10},
    		xticklabel style = {align=center, font=\small, rotate=300, anchor=west},
            xticklabels from table={\csvdata}{technique},
    		xtick style = {draw=none}, 
    		ylabel = {Cycle Time (seconds)},
            ymode=log
    	]
            \pgfplotsinvokeforeach{0,...,\TotalRows}
            {
              \addplot+[
              boxplot prepared from table={
                table=\csvdata,
                row=#1,
                lower whisker=min_mean,
                lower quartile=q1_mean,
                median=median_mean,
                average=mean_mean,
                upper quartile=q3_mean,
                upper whisker=max_mean
              },
              boxplot prepared,
              fill,
              draw=black
              ]
              coordinates {};
            }
    	\end{axis}
    \end{tikzpicture}

    \caption{Box plot charts of the cycle time distribution for each dataset and configuration of the proposed approach (the Y axis is in a logarithmic scale).}
    \label{fig:boxplot-cts}
\end{figure*}

Regarding the distribution of cycle times, \figurename~\ref{fig:boxplot-cts} depicts, for each dataset, the minimum, first quartile, median, average, third quartile, and maximum cycle times for the testing log (denoted as \textit{Ground Truth}), and for the logs simulated with each of the BPS models.
The results of the AC\_CRE and BPIC17 datasets show a similar pattern as the one seen by the RED distance.
SIMOD presents the farthest cycle time distribution from the test log, followed by $\mathit{Naive}$ and $\mathit{Naive^{TPE}}$, in that order.
Finally, $\mathit{EclAw+}$ and $\mathit{EclAw+^{TPE}}$ show the closest distribution to the test log, with small differences between them.
Also, similarly to the RED results, the differences between the ex-post and ex-ante versions of each approach are very small.
The results of the BPIC12 dataset behave slightly differently.
We can see how the BPS models that get closer to the ground truth are the BPS models enhanced with $\mathit{EclAw}$ and $\mathit{EclAw^{TPE}}$, especially the ex-ante configurations.
However, the other techniques are not as far as in the other datasets.

Finally, regarding the computational cost of the different variants of the approach, \tablename~\ref{tab:real-life-runtimes} shows the runtimes for the three real-life processes.
In most of the cases, the runtime difference between the ex-post and ex-ante configurations is negligible.
Regarding the proposals combined with the TPE optimizer phase, this addition causes a considerably high computational overhead, going from less than one minute to 20 in the BPIC12 dataset, and from three minutes to around one hour in the BPIC17 dataset.

\vspace{10pt}
\noindent\textbf{Discussion.}
The results of the experiments on the real-life logs show that the ex-ante configuration generally presents better results than the ex-post configuration.
The differences are small in $\mathit{Naive}$ and $\mathit{Naive^{TPE}}$, where both configurations present similar results in many scenarios.
On the other hand, in the case of $\mathit{EclAw+}$ and $\mathit{EclAw+^{TPE}}$, the improvement of the ex-ante configuration is considerably higher, especially in the BPIC12 and BPIC17 datasets.
However, as commented before, the difference between these two proposals is more conceptual than performance-related.
In the evaluated processes, the best option is to attribute the delay to the executed activity (ex-ante), but there might be other processes where attributing the delay to the enabling activity (ex-post) is the best option.

Regarding the differences between $\mathit{Naive}$ and $\mathit{EclAw+}$, the results show how the latter one outperforms the former in the three datasets.
As expected, the proposed eclipse-aware method is less sensible to eclipses, estimating extraneous delays that lead to enhanced BPS models that better reflect the temporal performance of the process.

Regarding the benefits of the TPE optimization version of the approaches, this process generally improves the results of the enhancement.
As explained before, the resource contention waiting time may partially eclipse the extraneous activity delays' durations, leading to shorter delays and, thus, shorter cycle times.
However, the computational cost of this improvement is very high.

Finally, the AC\_CRE dataset represents a good example for the consideration of extraneous delays in BPS.
In this dataset, the direct estimation of $\mathit{EclAw+}$ gets closer to the best results (slightly improved by $\mathit{EclAw+}^{TPE}$).
Some of the activities of this process have to be performed by resources external to the process in question, that is, resources that are not dedicated to this process, but that instead spend their time in activities elsewhere (e.g., in other processes).
These external resources have low participation in the process, and usually perform the activities close to their deadlines.
This causes extraneous delays.

\subsection{Threats to validity}

The evaluation reported above is potentially affected by the following threats to the validity.
First, regarding \textit{internal validity}, the experiments rely only on four simulated and three real-life processes.
The results could be different for other datasets.
Second, regarding \textit{construct validity}, we used a measure of goodness based on discretized distributions to analyze the temporal distribution of events within each trace, and compared the temporal performance w.r.t.\ the cycle time by using a set of summary statistics.
The results could be different with other measures, e.g., measures of distance between time series based on dynamic time warping, which provides an alternative framework for capturing rhythms in time series.
Finally, regarding \textit{ecological validity}, the evaluation compares the simulation results against the original log.
While this allows us to measure how well the simulation models replicate the as-is process, it does not allow us to assess the goodness of the simulation models in a what-if setting, i.e., predicting the performance of the process after a change.

%% file: conclusions.tex
\section{Conclusion and Future Work\label{sec:conclusions}}

This article presented a method to enhance a BPS model discovered from an event log to take into account waiting times that cannot be attributed either to resource contention or to resource unavailability, i.e., extraneous activity delays.
The method first estimates the extraneous activity delay between each pair of causally consecutive activity instances $\varepsilon_{1}$ and $\varepsilon_{2}$, by analyzing other activity instances between $\varepsilon_{1}$ and $\varepsilon_{2}$ performed by the resource who ultimately performed $\varepsilon_{2}$, as well as her/his/their working periods.
Then, for every activity $\alpha_{i}$ in the process, the proposed method fits a probability distribution function to model the extraneous delays associated with this activity, based on the estimated extraneous delays of all the causally consecutive activity instances related to $\alpha_{i}$.
The method then enhances the BPS model with (duration-based) timer events, to model these delays.

The article considers multiple variants of the proposed approach.
The first variant consists of a naive method ($\mathit{Naive}$) to estimate the extraneous delay of an activity instance by subtracting, to its start time, the latest of \textit{i)} the time when the activity became available, and \textit{ii)} the time when the resource who performed this activity instance last became available.
Two more variants are presented, consisting of a more sophisticated approach that takes into account other activity instances performed by the same resource since the completion of the causally preceding activity instance, as well as her/his/their unavailability periods, which may ``eclipse'' some of the extraneous delays.
One of these versions ($\mathit{EclAw}$) follows a conservative approach solely considering the evidence found in the event log, while the other ($\mathit{EclAw+}$) tries to reduce the error of the estimation by extrapolating the result of this evidence.
The article also considers variants of the approach where the delays are modeled via a timer after the source activity (ex-post) or, alternatively, via a timer before the target activity (ex-ante).
Finally, the article considers the possibility of further tuning the probability distributions of the extraneous delays by means of a hyperparameter optimization technique typically used to tune machine learning models (TPE).

The empirical evaluation shows that, by modeling extraneous waiting times, we obtain BPS models with higher accuracy than an approach that only captures waiting times due to resource contention and unavailability.
Further, the evaluation concludes that the best placement of timer events to model extraneous delays (ex-ante or ex-post) varies from one dataset to another, suggesting that both approaches should be considered on a case-by-case basis.
Regarding the proposed variants, the evaluation showed that the eclipse-aware proposals ($\mathit{EclAw}$ and $\mathit{EclAw+}$) enhance the accuracy of the BPS model in all cases, relative to $\mathit{Naive}$.
Among these two, $\mathit{EclAw+}$ presented more accurate results in all cases.
The hyperparameter tuning (TPE) can further enhance the accuracy of the eclipse-aware approach, but at a high computational cost.
In conclusion, the ex-ante $\mathit{EclAw+}$ version with TPE optimization ($\mathit{EclAw+^{TPE}}$) resulted in the best enhancement in all real-life datasets, followed by the ex-ante $\mathit{EclAw+}$ proposal (without TPE optimization).
Accordingly, we recommend the use of $\mathit{EclAw+^{TPE}}$ when the computational efficiency is not relevant, and $\mathit{EclAw+}$ for a better tradeoff in terms of performance and computational efficiency.




The proposed technique seeks to model delays not attributable to resource contention or unavailability.
These extraneous delays may stem from batch processing and prioritization policies~\cite{DBLP:conf/caise/LashkevichMCSD23}, or from a variety of resource behaviors, including fatigue effects, procrastination, stress, and workload effects (e.g., the Yerkes-Dodson Law of Arousal)~\cite{DBLP:conf/caise/NakatumbaWA12}.
A direction for future work is to further extend the proposed approach in order to accurately diagnose the different sources of extraneous delays.

Another limitation of the proposed approach is that it assumes that the start time of a case is equal to the start time of the first activity instance in a case.
In other words, it assumes that the first activity instance in each case does not have any waiting time.
Another avenue for future work is to extend the proposed approach to analyze the waiting times preceding the first activity in each case, for example by adapting the technique proposed in~\cite{DBLP:conf/bpm/MartinDC15,DBLP:conf/icpm/BerkenstadtGSSW20}.

Finally, the proposed approach models the extraneous delays as timer events associated with the preceding (ex-post) or the succeeding (ex-ante) activity.
We foresee that attributing the delay to both activities (associated with the causally consecutive pair) instead of only one of them might improve the accuracy of the proposed techniques.

In a similar vein, as stated in Section~\ref{subsec:limitation}, some delays may be dependent not on the (next) activity, but rather on the resource who performs this activity. The proposed approach indirectly captures such delays, but only in an aggregate manner, across all resources who perform the activity. Thus, another avenue for future work is to extend the proposed approach by explicitly capturing resource-dependent delays, or alternatively, by combining it with an approach that captures resource availability in a probabilistic manner~\cite{LopezPintado2023}, and which can therefore capture sporadic resource unavailability.


The approach inserts one-off duration timers, i.e., timers that, when enabled, delay the execution of the next activity by a certain duration.
Another direction for future work is to extend the approach with other types of timer events (e.g., periodic or interrupting events).




%% file: acknowledgements.tex
\paragraph{Acknowledgments}
Research funded by the European Research Council (PIX Project).